Article

# Unified Equation of State for Neutron Stars Based on the Gogny Interaction

Xavier Viñas [1],*, Claudia Gonzalez-Boquera [1], Mario Centelles [1], Chiranjib Mondal [2] and Luis M. Robledo [3,4]

1. Departament de Física Quàntica i Astrofísica and Institut de Ciències del Cosmos (ICCUB), Facultat de Física, Universitat de Barcelona, Martí i Franquès 1, E-08028 Barcelona, Spain; claugb@fqa.ub.edu (C.G.-B.); mario@fqa.ub.edu (M.C.)
2. Laboratoire de Physique Corpusculaire Caen, 6 Boulevard Marchal Juin, 14000 Caen cedex, France; mondal@lpccaen.in2p3.fr
3. Departamento de Física Teórica and CIAFF, Universidad Autónoma de Madrid, E-28049 Madrid, Spain; luis.robledo@uam.es
4. Center for Computational Simulation, Universidad Politécnica de Madrid. Campus de Montegancedo, Boadilla del Monte, 28660-Madrid, Spain
* Correspondence: xavier@fqa.ub.edu

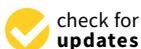

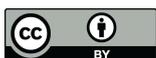

**Abstract:** The effective Gogny interactions of the D1 family were established by D. Gogny more than forty years ago with the aim to describe simultaneously the mean field and the pairing field corresponding to the nuclear interaction. The most popular Gogny parametrizations, namely D1S, D1N and D1M, describe accurately the ground-state properties of spherical and deformed finite nuclei all across the mass table obtained with Hartree–Fock–Bogoliubov (HFB) calculations. However, these forces produce a rather soft equation of state (EoS) in neutron matter, which leads to predict maximum masses of neutron stars well below the observed value of two solar masses. To remove this limitation, we built new Gogny parametrizations by modifying the density dependence of the symmetry energy predicted by the force in such a way that they can be applied to the neutron star domain and can also reproduce the properties of finite nuclei as good as their predecessors. These new parametrizations allow us to obtain stiffer EoS's based on the Gogny interactions, which predict maximum masses of neutron stars around two solar masses. Moreover, other global properties of the star, such as the moment of inertia and the tidal deformability, are in harmony with those obtained with other well tested EoSs based on the SLy4 Skyrme force or the Barcelona–Catania–Paris–Madrid (BCPM) energy density functional. Properties of the core-crust transition predicted by these Gogny EoSs are also analyzed. Using these new Gogny forces, the EoS in the inner crust is obtained with the Wigner–Seitz approximation in the Variational Wigner–Kirkwood approach along with the Strutinsky integral method, which allows one to estimate in a perturbative way the proton shell and pairing corrections. For the outer crust, the EoS is determined basically by the nuclear masses, which are taken from the experiments, wherever they are available, or by HFB calculations performed with these new forces if the experimental masses are not known.

## 1. Introduction

The standard Gogny interactions of the D1 family [1] consist of a finite-range part, which is modeled by two Gaussian form-factors including all the possible spin and isospin exchange terms, a zero-range density dependent term, which simulates the effect of the three-body forces, and a spin-orbit force, which is also of zero-range as in the case of Skyrme forces. Large-scale Hartree–Fock–Bogoliubov (HFB) calculations performed in a harmonic oscillator basis with the D1S parametrization [2,3] reveal that there is a systematic drift in the binding energy of neutron-rich nuclei (see [4] for more details). To overcome this deficiency, new parametrizations of the Gogny interaction, namely D1N [5] and D1M [6] were proposed. Unlike the D1S and D1N forces, whose parameters were obtained following the fitting protocol established in Ref. [1], the parameters of the D1M interaction were obtained by minimizing the energy *rms* deviation of 2149 measured nuclear masses of the AME2003 evaluation [7]. It is worthwhile to mention that, in the calibration of the D1N and D1M forces, in order to improve the description of neutron-rich nuclei, it was imposed that these interactions would follow the trend of the microscopic neutron matter EoS of Friedman and Pandharipande [8]. The D1M force reproduces the experimental nuclear masses of 2149 nuclei with an energy *rms* deviation of 798 keV. As an example, we display in the right panel of Figure 1 the binding energy differences between theoretical, computed with the D1M force at HFB level [9,10], and experimental binding energies, taken from the 2012 mass evaluation [11], of 620 even–even spherical and deformed nuclei. The theoretical binding energies include the HFB contribution and the rotational energy correction. However, the quadrupole zero point energy correction, which was included in the original fit, is approximated by a constant shift in the energy. We see that these differences are scattered around zero and do not show any energy drift for large neutron numbers. In the left panel of the same Figure, we display the same differences but computed with the D1S force. In this case the previously mentioned drift of binding energies can be clearly appreciated.

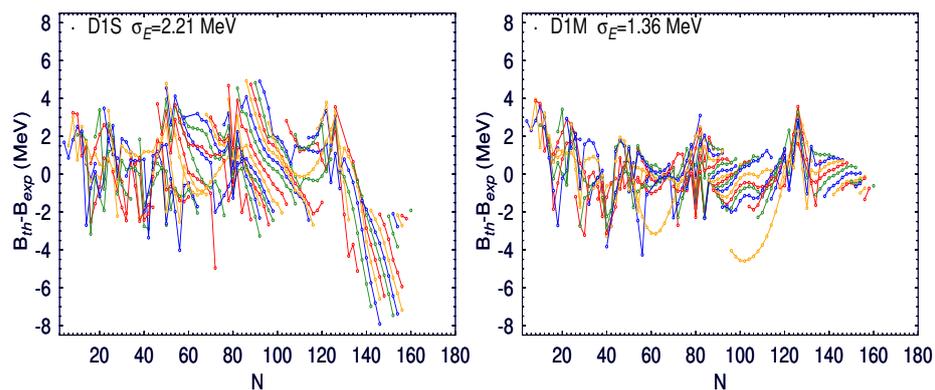

**Figure 1.** Differences between the computed and the experimental binding energies of 620 even–even nuclei. Theoretical calculations are performed with the Gogny D1S (left panel) and D1M (right panel) interactions. The experimental values are taken from [11].

However, the use of Gogny interactions in the neutron star (NS) domain does not work so well. In recent years it has been shown [12,13] that the most successful Gogny parametrizations, namely D1S, D1N and D1M, fall short in predicting a maximum NS mass of two solar masses ($M_\odot$), as required by some well contrasted astronomical observations [14–17]. A new extension of the Gogny force with a finite–range density–dependent term has been recently postulated [18]. This interaction, denoted as D2, has not been used much in finite nuclei calculations due to the complexity introduced by the finite range of the density–dependent term, but its EoS is able to reproduce the correct limit for the NS masses [19,20]. The structure of a standard NS composed by neutrons, protons and leptons (electrons and muons) in charge and in $\beta$–equilibrium is driven by its EoS, which allows the expression of the total pressure $P$ of the system to be written as a function of the baryonic density $\rho$. The EoS is the essential input needed to solve the Tolman–Oppenheimer–Volkov (TOV) equations, whose solution provides the mass–radius relationship of the NS. Throughout this work we consider that the NS is non-rotating, cold and locally charge neutral and in absolute thermodynamic equilibrium. This is a reasonable picture for an NS that was created a long time ago and had enough time to cool down.

In the uniform core of the star, the total pressure is given by the sum of the baryonic ($P_b$) and leptonic ($P_l$) contributions:

$$P = P_b + \sum_l P_l = \rho^2 \frac{\partial E_b}{\partial \rho} + \sum_l \rho_l^2 \frac{\partial E_l}{\partial \rho_l}, \qquad (1)$$

where $l = e, \mu$. In (1) $E_b$ and $E_l$ are the baryon and lepton energies per particle and $\rho = \rho_n + \rho_p$ is the total baryon density with $\rho_n$ and $\rho_p$ being the neutron and proton densities, respectively. The lepton densities $\rho_l$, owing to the charge equilibrium, are related to the proton density by $\rho_p = \rho_e + \rho_\mu$, where $\rho_e$ and $\rho_\mu$ are the electron and muon densities. Changing from the neutron and proton densities to the total density $\rho$ and to the isospin asymmetry $\delta = (\rho_n - \rho_p)/\rho$, each contribution to the total pressure (1) can also be written as

$$\begin{aligned} P_b &= \mu_n \rho_n + \mu_p \rho_p - \mathcal{H}_b(\rho, \delta) \\ P_l &= \mu_l \rho_l - \mathcal{H}_l(\rho_l), \end{aligned} \qquad (2)$$

where $\mathcal{H}_b$ and $\mathcal{H}_l$ are the baryonic ($b$) and leptonic ($l = e, \mu$) energy densities and $\mu_n, \mu_p, \mu_e$ and $\mu_\mu$ are the neutron, proton, electron and muon chemical potentials, respectively, which are defined as

$$\mu_n = \frac{\partial \mathcal{H}_b}{\partial \rho_n}; \quad \mu_p = \frac{\partial \mathcal{H}_b}{\partial \rho_p}; \quad \mu_e = \frac{\partial \mathcal{H}_e}{\partial \rho_e}; \quad \mu_\mu = \frac{\partial \mathcal{H}_\mu}{\partial \rho_\mu}. \qquad (3)$$

In stable neutron star matter (NSM) the direct Urca processes

$$n \to p + l + \bar{\nu}_l \qquad \text{and} \qquad p + l \to n + \nu_l \qquad (4)$$

take place simultaneously. Assuming that the neutrinos eventually leave the star, the $\beta$-equilibrium condition leads to

$$\mu_n - \mu_p = \mu_e = \mu_\mu. \qquad (5)$$

The EoSs for NSM in logarithmic scale as a function of the baryonic density computed for some of the Gogny interactions used in this work and obtained previously in Refs. [12,13] are displayed in Figure 2 together with the EoS provided by the BCPM energy density functional [21], which we will use here as a benchmark, as well as the EoS obtained using the SLy4 [22] and BSk22 [23] Skyrme forces. The BCPM EoS, derived in the framework of the microscopic Brueckner–Bethe–Goldstone theory (see [21] and references therein), is in very good agreement with the EoS provided by the SLy4 force [22], which was specifically built for astrophysical calculations. We can also see that the EoS corresponding to the BSk22 Skyrme force obtained by the Brussels–Montreal group and reported in Ref. [23] (also see Ref. [24]) is stiffer than the EoSs computed with the SLy4 Skyrme force and the BCPM energy density functional. From this Figure we can see that the EoSs obtained with the D1N and D1M forces show an increasing trend with growing baryon density but softer than the behavior exhibited by the BCPM EoS. We can also see that the EoS for NSM calculated with the D1S force reaches a maximum value at around twice the normal saturation density and decreases for larger densities. As a consequence of this anomalous behavior, the TOV equations cannot be solved in the D1S case, which implies that the D1S interaction is not well suited for astrophysical calculations. The shaded area in Figure 2 depicts the region in the $P$-$\rho$ plane consistent with the experimental collective flow data in Heavy-Ion Collisions (HIC) [25]. From this Figure we can see that none of the EoSs computed with the standard Gogny interactions are able to clearly pass through the region constrained by the collective flow in HIC.

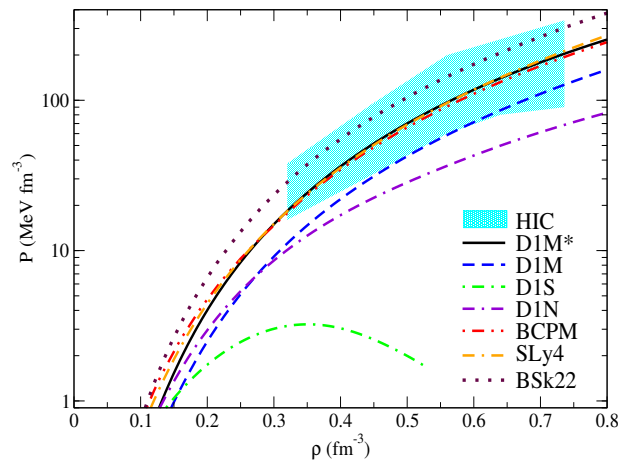

**Figure 2.** Equation of state (total pressure in logarithmic scale against baryon density) for neutron star matter computed with the D1M*, D1M, D1N and D1S Gogny interactions, with the BCPM energy density functional and with the SLy4 and BSk22 Skyrme forces. Constraints coming from collective flow in heavy-ion collisions are also included [25].

The baryonic part of the EoS is basically driven by the energy density of highly asymmetric nuclear matter (ANM) $\mathcal{H}_b(\rho,\delta)$, where the isospin asymmetry $\delta$ takes values around 0.9. To characterize this energy density, which is close to the pure neutron matter, it is extremely useful to introduce the symmetry energy, which can be understood as the energy cost to convert all protons into neutrons in symmetric nuclear matter. The energy per particle $E_b(\rho,\delta)=\mathcal{H}_b(\rho,\delta)/\rho$ in ANM can be written as a Taylor expansion with respect to the isospin asymmetry around $\delta=0$:

$$E_b(\rho,\delta) = E_b(\rho,\delta=0) + \sum_{k=1}^{\infty} E_{sym,2k}(\rho)\delta^{2k}, \qquad (6)$$

where we have assumed the charge symmetry of the strong interaction, which implies that only even powers of $\delta$ appear in (6). The first term of the expansion, $E_b(\rho,\delta=0)$ is the energy per baryon in symmetric nuclear matter and the coefficients of the Taylor expansion are given by:

$$E_{sym,2k} = \frac{1}{(2k)!}\frac{\partial^{2k}E_b(\rho,\delta)}{\partial\delta^{2k}}\bigg|_{\delta=0}. \qquad (7)$$

The symmetry energy coefficient $E_{sym}$ is usually defined as the second-order coefficient in the expansion (6), i.e., $E_{sym} \equiv E_{sym,2}$. In many cases the energy per particle in ANM is well approximated taking only the quadratic term in the expansion (6), that is,

$$E_b(\rho,\delta) = E_b(\rho,\delta=0) + E_{sym}(\rho)\delta^2. \qquad (8)$$

Therefore, it is also possible to define the symmetry energy as the difference between the energy per particle in pure neutron matter and in symmetric nuclear matter,

$$E'_{sym} = E_b(\rho,\delta=1) - E_b(\rho,\delta=0). \qquad (9)$$

Taking into account (6), it is clear that the definition (9) corresponds to the whole sum of the coefficients $E_{sym,2k}$. The difference between both definitions of the symmetry energy depends on the importance of the contribution of the terms higher than the quadratic one in the expansion (6). A detailed discussion about the higher-order symmetry energy contributions in the case of Gogny interactions can be found in Refs. [13,20]. In Figure 3 we display the symmetry energy, defined as Equation (7) with $k=1$, as a function of the baryonic density computed with different Gogny forces available in the literature and taken from Refs. [12,13]. In the same Figure we also show the symmetry energy constraints extracted from the isobaric analog states (IAS) and from IAS combined with neutron skins [26], the constraints from the electric dipole polarizability $\alpha_D$ in $^{208}$Pb [27] and from transport simulations in heavy-ion collisions in Sn isotopes [28].

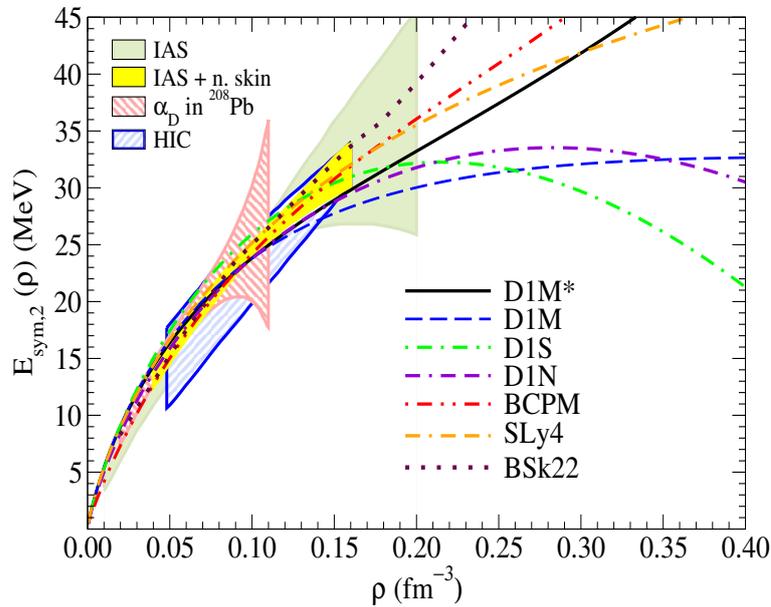

**Figure 3.** Symmetry energy, defined as Equation (7) with $k = 1$ against the baryon density predicted by the D1M*, D1M, D1S and D1N Gogny interactions, the BCPM energy density functional and the SLy4 and BSk22 Skyrme forces. Some constraints coming from isobaric analog states (IAS) (green), from IAS plus neutron skins (IAS + n.skin) (yellow), electric dipole polarizability $\alpha_D$ in $^{208}$Pb ($\alpha_D$ in $^{208}$Pb) (dashed red) and heavy-ion collisions (dashed blue) are also included [26–28].

From Figure 3 we can see that, below the saturation density, the symmetry energy behaves in a very similar way for all the considered forces taking values around 30 MeV at saturation. This is due to the fact that in this region the symmetry energy is well constrained by the nuclear masses to which the parameters of the different effective interactions have been fitted (see Refs. [29–31] for a review about the range of the symmetry energy obtained from different constraints). Above the saturation density, the symmetry energy predicted by the different interactions differ more among them. For example, we can see that the symmetry energy computed with the D1S and D1N parametrizations reaches maximum values of 30–40 MeV, and then decrease with increasing density until vanishing around 3–4 times the saturation density, where the isospin instability starts. In the case of the D1M force the symmetry energy also reaches a maximum value, which remains practically constant in the whole density range needed to solve the TOV equations. From the same Figure 3 we also observe that the symmetry energy computed with the BCPM energy density functional shows a different trend, growing with increasing density. The symmetry energy computed with the SLy4 and BSk22 Skyrme forces, which provide realistic EoSs, also shows an increasing trend with growing density, BSk22 being stiffer and SLy4 softer in the high-density domain above 0.20 fm$^{-3}$. These results show that the behavior of the symmetry energy as a function of the density above the saturation is crucial for describing properly the EoS of neutron-rich matter in the high-density regime, which, in turn, is the most relevant input for the study of many NS properties.

An important feature of the symmetry energy is its density content calculated at saturation density. This quantity is usually characterized by the slope of the symmetry energy $L$, which is defined as

$$L = 3\rho_0 \left.\frac{\partial E_{sym}(\rho)}{\partial \rho}\right|_{\rho_0}. \tag{10}$$

The slope parameter is connected with different properties of finite nuclei, as for example the neutron skin thickness in heavy nuclei such as $^{208}$Pb (see [32–35] and references therein). The numerical values of the slope parameter $L$ predicted by different models span a very large range between 10 and 120 MeV, pointing out that this quantity is poorly constrained by the available experimental data. A compilation of possible $L$ values extracted from different laboratory experiments and astronomical observations can be found in Refs. [19,36,37]. From the theoretical side, some recent microscopic calculations have estimated the slope parameter in the ranges $L$ = 43.8–48.6 MeV [38], $L$= 20–65 MeV [39] and $L$ = 45–70 MeV [40]. The values of the slope parameter predicted by the standard Gogny forces of the D1 family are relatively small, $L$ = 22.43 MeV (D1S), $L$ = 24.83 MeV (D1M) and $L$ = 33.58 MeV (D1N) [12]. These values, which are clearly smaller than the value $L$ = 52.96 MeV predicted by the BCPM energy density functional and those of the SLy4 and BSk22 Skyrme forces, clearly explain the soft behavior of the symmetry energy displayed in Figure 3 and consequently the softness of the EoS in NS matter predicted by such forces (see Figure 2). In Figure 4 we display some bounds of the symmetry energy at saturation $E_{sym}(\rho_0)$ and its slope $L$ provided by recent laboratory data, astronomical observations and *ab initio* calculations using chiral interactions [30,36,41,42]. We see that the symmetry energy and its slope predicted by the Gogny forces D1M and D1N lie outside the constrained region in the $E_{sym}(\rho_0)$-$L$ plane, while the point corresponding to the D1S interaction is at the lower edge of the region estimated from the measured electric dipole polarizability in $^{68}$Ni, $^{120}$Sn and $^{208}$Pb [41].

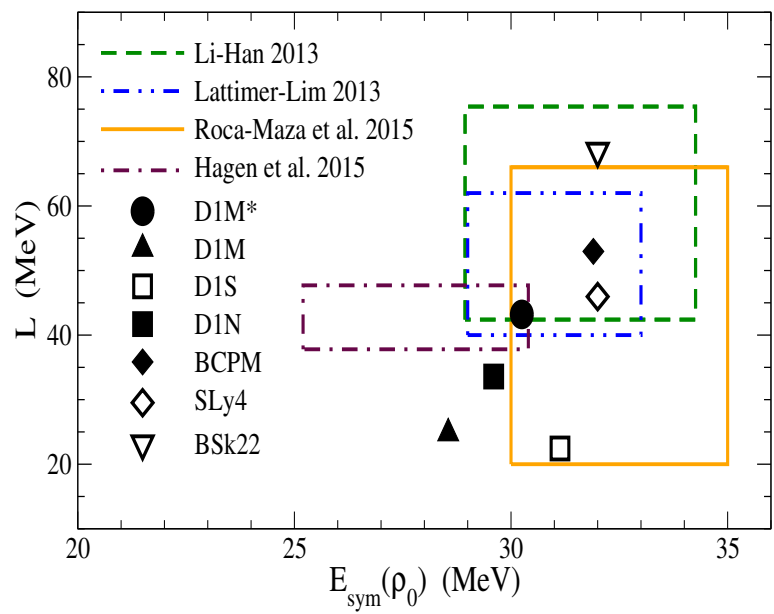

**Figure 4.** Slope of the symmetry energy $L$ against the symmetry energy at saturation density for some Gogny interactions, the BCPM energy density functional and the SLy4 and BSk22 Skyrme forces. We have included some constraints extracted from the literature [30,36,41,42].

From this discussion it is clear that the standard Gogny interactions of the D1 family are not well suited for applications in the NS domain. To overcome this situation we designed some parametrizations of the Gogny type of forces starting from the D1M interaction [19,20,43,44] aimed to predict a maximum mass in NS of $2M_\odot$ without losing its ability to describe finite nuclei with a quality similar to those found using the D1M force. The purpose of this paper is to review those new parametrizations and compare them with previous results. The paper is organized as follows. In the second section we describe the method used to fit these new Gogny parametrizations, namely D1M$^*$ and D1M$^{**}$. In the third section we describe how the EoS in the inner and outer crust using the D1M$^*$ interaction is obtained. In the same section the study of the core–crust transition using the thermodynamical and dynamical methods is briefly summarized. The fourth section is devoted to discussing some global NS properties such as the mass–radius relation, the moment of inertia, its crustal properties and the tidal deformability estimated with the new Gogny interaction D1M*. We also compare in this section the D1M* results with the predictions provided by other different models. Finally, our conclusions are presented in the last section.

## 2. Gogny Interactions Adapted for Astrophysical Calculations

The standard Gogny interaction of the D1 family consists of a finite range term, which is modeled by two form factors of Gaussian type and includes all possible spin and isospin exchange terms, plus a zero-range density-dependent contribution. To describe finite nuclei, a spin–orbit interaction—which is zero-range like in the case of Skyrme forces—is also added. With all these ingredients the Gogny interaction reads:

$$
\begin{aligned}
V(\mathbf{r}_1, \mathbf{r}_2) &= \sum_{i=1,2} \left( W_i + B_i P_\sigma - H_i P_\tau - M_i P_\sigma P_\tau \right) e^{-\frac{r^2}{\mu_i^2}} + t_3(1 + x_3 P_\sigma)\rho^\alpha(\mathbf{R})\delta(\mathbf{r}) \\
&\quad + i W_{LS}(\sigma_1 + \sigma_2)\{\mathbf{k}' \times \delta(\mathbf{r})\mathbf{k}\},
\end{aligned}
\tag{11}
$$

where $\mathbf{r}$ and $\mathbf{R}$ are the relative and the center of mass coordinates of the two nucleons, and $\mu_1 \simeq$ 0.5–0.7 fm and $\mu_2 \simeq$ 1.2 fm are the ranges of the two Gaussian form factors, which simulate the short- and long-range components of the force, respectively. The Skyrme-type $t_3$ and $x_3$ parameters control the density dependent part of the force.

To determine the parameters of the new Gogny interactions, denoted D1M* and D1M**, we start from the D1M force and modify the parameters of the finite-range part of the interaction, which are the ones that control the stiffness of the symmetry energy, keeping the binding energy and charge radius of finite nuclei predicted by these interactions as close as possible to the values obtained with the original D1M force. This way of proceeding is similar to the one used with some Skyrme forces and RMF parametrizations, such as SAMi-J [45], KDE0-J [46] or FSU-TAMU [47,48].

Therefore, we readjust the eight parameters $W_i$, $B_i$, $H_i$ and $M_i$ ($i = 1, 2$) of the finite-range part of the Gogny interaction. The ranges of the two Gaussian form factors and the zero-range part of the force are kept fixed to the original values of D1M. The open parameters are constrained by imposing in symmetric nuclear matter the same values of the saturation density, energy per particle, incompressibility modulus and effective mass as the ones predicted by the original D1M force. It has been claimed in earlier literature that finite nuclei energies constrain the symmetry energy at a subsaturation density of about 0.1 fm$^{-3}$ better than at saturation density [32,49]. Hence, we impose that the symmetry energy of the modified interaction at this particular density also equals the corresponding value provided by the D1M force. In order to preserve the pairing properties of D1M we also require that, in the new force the combinations $W_i - B_i - H_i + M_i$ ($i$ = 1,2), which govern the strength of the pairing interaction, take the same value as in the original D1M force. There is still an open parameter, which we chose to be $B_1$. This parameter is used to modify the slope of the symmetry energy at saturation $L$, which in turn determines the maximum mass of the neutron star. We adjust this parameter $B_1$ in such a way that the maximum mass computed with the new parametrizations of the Gogny force are $2M_\odot$ (D1M*) and $1.91M_\odot$ (D1M**). Finally, we perform a fine tuning of the strength $t_3$ of the density-dependent term of the interaction in order to optimize the description of the masses of finite nuclei. To this end we compute the energies of 620 spherical and deformed even–even nuclei of the AME2012 database[11] at HFB level using the HFBaxial code [9]. As it is customary with Gogny forces, we carry out the HFB calculations in a harmonic oscillator basis. The parameters and size of the basis are chosen as to optimize the binding energies for each value of mass number $A$. An approximate second-order gradient is used to solve with confidence the HFB equations [10]. It has been known for a long time that some Skyrme parametrizations present numerical instabilities when the finite-nuclei calculations are performed on a mesh in a coordinate space, see e.g., [50] and references therein. It has been recently shown that the Gogny parameter sets may also display finite-size instabilities [51] that lead to diverging results in the coordinate-space calculations of finite nuclei [51,52]. This is the case of the D1N and D1M* forces [51,52] and, to a lesser extent, of D1M [52]. Therefore, the HFB calculations of finite nuclei with the new parameter set D1M* are to be performed in a harmonic oscillator basis [19,52]. The numerical values of the parameters of the new forces D1M* and D1M** were reported in Refs. [19,20,43]. For the sake of completeness, we collect them also here in Table 1, along with the parameters of D1M. In Table 2 we report the nuclear matter properties predicted by the D1M* and D1M** forces, as well as by the BCPM energy density functional, which is used in this work as a benchmark for comparison with the results provided by the new Gogny parametrizations D1M* and D1M**.

From Table 1, we observe that the finite-range parameters $W_i$, $B_i$, $H_i$ and $M_i$ of the modified D1M* and D1M** forces are larger in absolute value than the ones in the original D1M interaction. However, as can be seen in Table 2, the saturation properties of symmetric nuclear matter (namely, the saturation density $\rho_0$, the energy per particle $E_0$ at saturation, the incompressibility $K_0$, and the effective mass $m^*/m$) and the symmetry energy at a density 0.1 fm$^{-3}$, predicted by the D1M** interaction coincide with the values computed with the D1M force as a consequence of the fitting protocol used to obtain the parameters of the modified forces. In the case of the D1M* force we also slightly changed the $t_3$ parameter by an amount of 1 MeV to improve the finite nuclei description with this interaction. As a consequence of this small change in $t_3$, the symmetric nuclear matter properties involved in the reparametrization changes slightly compared to the corresponding values predicted by the D1M force, as can be seen in Table 2. The properties that differ significantly between the new parametrizations and D1M are the symmetry energy at saturation density ($E_{sym}(\rho_0)$) and, visibly, the density dependence of the symmetry energy, which governs the isovector part of the interaction. The latter is quantified by the slope parameter $L$, which varies from a value $L$ = 24.84 MeV in the original D1M force to $L$ = 43.18 MeV for D1M* and to $L$ = 33.91 MeV for D1M**, as required to obtain a stiffer EoS in NS matter, which in turn allows predictions of the maximum mass of $2M_\odot$ and $1.91M_\odot$, respectively.

**Table 1.** Parameters of the D1M, D1M* and D1M** Gogny forces. The coefficients $W_i$, $B_i$, $H_i$ and $M_i$ are given in MeV, $\mu_i$ in fm and $t_3$ in MeV fm$^4$. The values of the other parameters of the modified interactions are the same as in the D1M force (namely, $x_3$=1, $\alpha$=1/3 and $W_{LS}$ = 115.36 MeV fm$^5$).

| D1M   | $W_i$       | $B_i$      | $H_i$       | $M_i$      | $\mu_i$ |
|-------|-------------|------------|-------------|------------|---------|
| i=1   | −12797.57   | 14048.85   | −15144.43   | 11963.81   | 0.50    |
| i=2   | 490.95      | −752.27    | 675.12      | −693.57    | 1.00    |
|       | $t_3$       | $x_3$      | $\alpha$    | $W_{LS}$   |         |
|       | 1562.22     | 1          | 1/3         | 115.36     |         |
| D1M*  | $W_i$       | $B_i$      | $H_i$       | $M_i$      | $\mu_i$ |
| i=1   | −17242.0144 | 19604.4056 | −20699.9856 | 16408.3344 | 0.50    |
| i=2   | 675.3860    | −982.8150  | 905.6650    | −878.0060  | 1.00    |
|       | $t_3$       | $x_3$      | $\alpha$    | $W_{LS}$   |         |
|       | 1561.22     | 1          | 1/3         | 115.36     |         |
| D1M** | $W_i$       | $B_i$      | $H_i$       | $M_i$      | $\mu_i$ |
| i=1   | −15019.7922 | 16826.6278 | −17922.2078 | 14186.1122 | 0.50    |
| i=2   | 583.1680    | −867.5425  | 790.3925    | −785.7880  | 1.00    |
|       | $t_3$       | $x_3$      | $\alpha$    | $W_{LS}$   |         |
|       | 1562.22     | 1          | 1/3         | 115.36     |         |

Let us now briefly discuss the main properties and predictions of these modified Gogny forces. As can be seen from Figure 3, the symmetry energy as a function of the baryon density obtained using D1M* shows a different behavior compared to the one exhibited by the standard Gogny interactions D1S, D1N and D1M. Above saturation the symmetry energy computed with D1M* increases with growing density and takes values close to the ones predicted by the BCPM energy density functional. As a consequence of this behavior, in the high–density domain the EoS predicted by the D1M* interaction follows closely the trend of the BCPM EoS, passing nicely through the region of the $P - \rho$ plane constrained by the experimental data of the heavy-ion collisions as can be seen in Figure 2. Finally let us point out that the representative points of the D1M* force lie within the region of the $E_{sym}(\rho_0)$-L plane constrained by the majority of the experimental data, as is seen in Figure 4. In order to check the ability of the D1M* force to describe finite nuclei, we plot in Figure 5 the differences between the binding energies of a set of 620 even–even nuclei computed with this new force and with the original D1M interaction along different isotopic chains covering the whole nuclear chart. We see that these differences are actually very small, lying within a window of ± 3 MeV for all the computed nuclei. As a general trend, the binding energy predicted by D1M* is larger than the one provided by D1M for neutron deficient nuclei of the isotopic chains and the opposite happens for neutron rich nuclei of the chain.

**Table 2.** Nuclear matter properties predicted by the D1M*, D1M** and D1M Gogny interactions and by the BCPM energy density functional.

|  | $\rho_0$ (fm$^{-3}$) | $E_0$ (MeV) | $K_0$ (MeV) | $m^*/m$ | $E_{sym}(\rho_0)$ (MeV) | $E_{sym}(0.1)$ (MeV) | $L$ (MeV) |
|---|---|---|---|---|---|---|---|
| D1M   | 0.1647 | −16.02 | 224.98 | 0.746 | 28.55 | 23.80 | 24.83 |
| D1M*  | 0.1650 | −16.06 | 225.38 | 0.746 | 30.25 | 23.82 | 43.18 |
| D1M** | 0.1647 | −16.02 | 224.98 | 0.746 | 29.37 | 23.80 | 33.91 |
| BCPM  | 0.1600 | −16.00 | 213.75 | 1.000 | 31.92 | 24.20 | 52.96 |

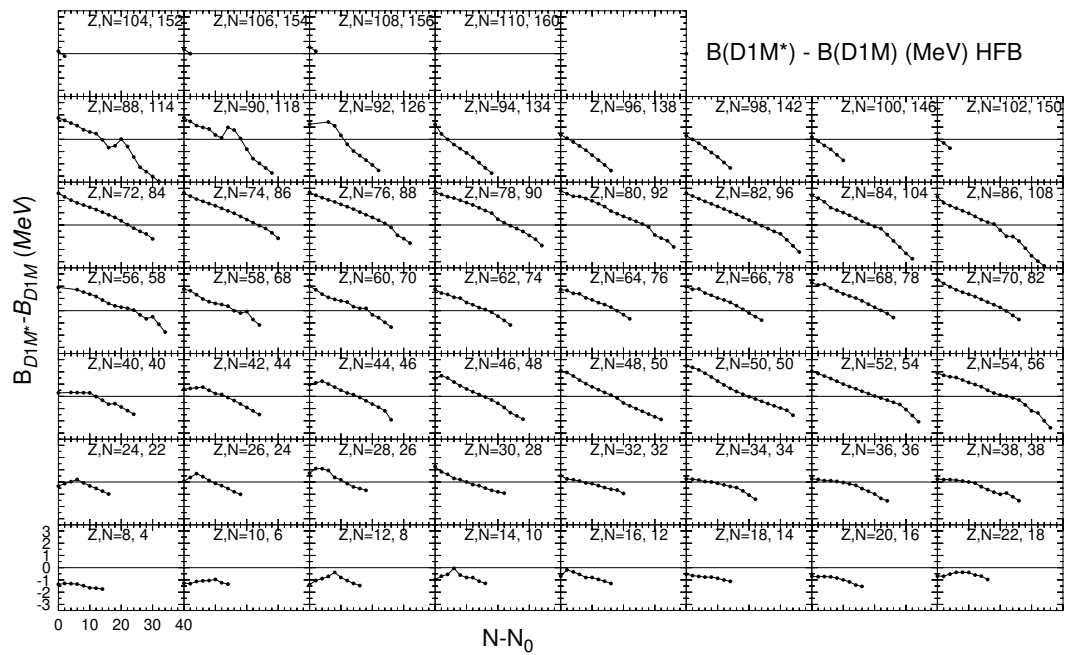

**Figure 5.** Difference between the binding energies provided by the D1M* and D1M force $\Delta B$ (in MeV) plotted as a function of the shifted neutron number $N$-$N_0$ for isotopic chains covering the periodic table. The values of the atomic number Z and neutron reference number $N_0$ are given in each panel. The vertical scale covers from +3.5 MeV to −3.5 MeV, with long ticks every MeV and short ticks every half MeV. The horizontal line in each panel at $\Delta B$=0 is plotted guide the eye.

## 3. Neutron Star Crust with Gogny Forces

The outer layer of an NS encircling the homogeneous core is denoted as "crust". It is further subdivided into two or three layers depending on its composition. At the surface of the star, namely the "outer crust", the matter is distributed in a lattice of neutron-rich nuclei immersed in an electron gas. After a certain density $\sim$0.003 fm$^{-3}$ going towards the center of the star, neutrons start to drip from the nuclei forming a background neutron gas but keeping a lattice structure of nuclear clusters. This region is denoted as the NS "inner crust". At a density $\sim$0.08 fm$^{-3}$, also known as the "crust–core transition density", the inner crust dissolves into an homogeneous core, sometimes with pasta phases in the transition region. As these complicated structures incorporate in-medium many-body effects, a full quantum mechanical treatment of the inner-crust is very difficult and computationally expensive. Nevertheless, there exist some calculations of the EoS in this region of NSs of different degrees of sophistication available in the literature (see for example [21,53–55] for references and a more detailed discussion on this topic). Simplified calculations based on the Thomas–Fermi (TF) approximation or its extended versions are often employed to obtain the EoS of the neutron star crust with different interactions (see [21,53–57] and references quoted therein). Even though global properties like the mass or the radius are not heavily influenced by the crustal properties of the NS, pulsar glitches, quasi-periodic oscillations in soft $\gamma$-ray repeaters or thermal relaxations in soft *X*-ray transients are strongly influenced by the crustal composition of the NS (see for example [54,55,58] and references quoted therein). The crust also might be one of the possible places where the *r*-process nucleosynthesis occurs during the NS–NS or NS–Black Hole merger events [59–61].

We have organized the description of the crust in this section as follows. In the first subsection we outline the variational Wigner–Kirkwood (VWK) method for describing finite nuclei. After that we describe the restoration of quantum effects like the shell correction with the Strutinsky integral method and the residual pairing correction with state dependent Bardeen–Cooper–Schrieffer (BCS) calculations. In the next subsection we compute with Gogny interactions the structure of the outer crust of a cold, non-accreting star. These calculations are performed within the so-called Wigner–Seitz (WS) approximation, which assumes that the space can be described by non-interacting electrically neutral cells, each one containing a single nuclear cluster embedded in electron (outer crust) or electron and neutron (inner crust) gases. In the inner crust, we restrict ourselves to spherically symmetric nuclear clusters disregarding pasta phases for the sake of simplicity. The results obtained with different Gogny forces are also displayed in the relevant subsections. At the end, we discuss the crust–core transition obtained with different Gogny interactions.

### 3.1. Variational Wigner-Kirkwood Method in Finite Nuclei

Semiclassical estimates of the binding energy of nuclei throughout the whole nuclear chart have been used since the Bethe–Weizsäcker mass formula was proposed [62,63]. The smooth part of the energy can be estimated by considering a Fermi gas-like system of nucleons with different choice of interactions. Further, one can treat quantum shell corrections perturbatively on top of it, using the techniques established by Strutinsky [64]. The residual pairing energy can also be calculated perturbatively using the shell structure corresponding to the average mean-field. The smooth part of the binding energy, i.e., neglecting quantal effects, of a set of non-interacting fermions in an external potential, can easily be obtained using the Wigner–Kirkwood (WK) $\hbar$-expansion of the single particle partition function [65–69]. An important feature of this expansion is that the variational solution of the minimization of WK energy at each $\hbar$-order is simply the WK expansion of the density at the same order. This method of solving a variational equation by sorting order-by-order the $\hbar$-expansion is called the VWK theory, which is discussed in detail in Refs. [68,70,71]. A primary feature of this method is that one needs to calculate one less order in the density expansion to accurately calculate the energy in the next order. For example, a VWK prediction on the energy containing $\hbar^2$-order contribution only needs the information on the $\hbar^0$-order densities, i.e., the bare TF densities.

To calculate the smooth part of the energy with the VWK method using the Gogny interaction (11), we use in this work the extended TF density matrix [72], which allows us to obtain the kinetic and exchange energy densities up to $\hbar^2$ order as a functional of the particle densities of each type of nucleons [53,73]. Therefore we write the VWK energy as

$$\begin{aligned} E_{\text{VWK}} &= \int \mathcal{H} d\mathbf{R} = \int (\mathcal{H}_0 + \mathcal{H}_2) d\mathbf{R} \\ &= \int (\mathcal{H}_{kin,0} + \mathcal{H}_{dir} + \mathcal{H}_{exch,0} + \mathcal{H}_{zr} + \mathcal{H}_{Coul}) d\mathbf{R} \\ &\quad + \int (\mathcal{H}_{kin,2} + \mathcal{H}_{exch,2} + \mathcal{H}_{SO}) d\mathbf{R}, \end{aligned} \quad (12)$$

where we have decomposed the energy into TF (subindex 0) and $\hbar^2$ (subindex 2) terms. For a detailed derivation of the energy density in (12), the reader is referred to Refs. [53,73].

To find the density profiles, which in turn will allow one to determine the VWK energy, one should solve first the variational TF equations for each type of particles with respect to the TF densities $\rho_q (q = n, p)$,

$$\frac{\delta}{\delta \rho_q} \left[ E_{\text{VWK},0} - \mu_q \int \rho_q(\mathbf{R}) d\mathbf{R} \right] = 0, \quad (13)$$

where $\mu_q$ are the chemical potentials that ensure the right number of nucleons of each type. Using the solutions of Equation (13) in Equation (12), one can calculate the semiclassical energy up to $\hbar^2$-order in the VWK approach.

Instead of solving the set of Equation (13), we perform a restricted variational calculation by minimizing the TF part of the VWK energy Equation (12) using a trial density of the Fermi type for each type of particles,

$$\rho_q(r) = \frac{\rho_{0,q}}{1 + \exp\left(\frac{r - C_q}{a_q}\right)}, \quad (14)$$

where the radius $C_q$ and the diffuseness parameter $a_q$ of each trial density are the variational parameters and the strengths $\rho_{0,q}$ are fixed by normalizing the neutron and proton numbers. Finally, using these trial densities the $\hbar^2$ part of the VWK energy in Equation (12) is added perturbatively. This restricted minimization of the energy with parametrized neutron and proton densities has been applied successfully in many semiclassical calculations of the energy of finite nuclei using Skyrme interactions [69], the differences with the full variational calculation being very small [74].

### 3.2. Shell and Pairing Effects

Once the average smooth part of the energy is determined, we add perturbatively the quantum shell energy that is obtained using the so-called Strutinsky integral method [75,76]. In this approximation, the shell correction is estimated as the difference between the quantal energy and its semiclassical counterpart of a set of nucleons moving under the action of an external single-particle Hartree–Fock Hamiltonian (see Refs. [53,73] for more details) generated by the parametrized neutron and proton densities (14). The corresponding Schrödinger equations read,

$$h_q \phi_{i,q} = \left\{ -\nabla \frac{\hbar^2}{2\widetilde{m}_q^*(\mathbf{r})} \nabla + \widetilde{U}_q(\mathbf{r}) - i\widetilde{\mathbf{W}}_q(\mathbf{r})(\nabla \times \sigma) \right\} \phi_i = \widetilde{\epsilon}_{i,q} \phi_{i,q}. \quad (15)$$

It should be noticed that the local particle $\widetilde{\rho}_q$, kinetic energy $\widetilde{\tau}_q$ and spin $\widetilde{J}_q$ densities, which are used to calculate the effective mass $\widetilde{m}_q^*$, the mean-field $\widetilde{U}_q$ and the spin-orbit potential $\widetilde{\mathbf{W}}_q$ appearing in Equation (15), are obtained semi-classically by the restricted variational approach described above.

After the single-particle energies $\widetilde{\epsilon}_{i,q}$ are obtained by solving Equation (15), the shell correction energy for each type of particles is given by

$$E_q^{shell} = \sum_i \widetilde{\epsilon}_{i,q} - \int \left[\frac{\hbar^2}{2\widetilde{m}_q^*}\widetilde{\tau}_q + \widetilde{\rho}_q \widetilde{U}_q + \widetilde{\mathbf{J}}_q \cdot \widetilde{\mathbf{W}}_q\right] d\mathbf{R}. \tag{16}$$

These single-particle energies $\widetilde{\epsilon}_{i,q}$ can be further used to calculate perturbatively the residual neutron and proton pairing energy through a BCS pairing calculation as,

$$E_q^{pair} = -\frac{1}{4}\sum_{k,q} \frac{\Delta_{k,q}^2}{E_{k,q}}, \tag{17}$$

where $E_{k,q}$ and $\Delta_{k,q}$ are the quasiparticle energy and the gap in the state $k$ of the type of particles $q$, respectively. The quasi-particle energy in the state $k$ reads

$$E_{k,q} = \sqrt{(\widetilde{\epsilon}_{k,q} - \mu_q)^2 + \Delta_{k,q}^2}, \tag{18}$$

which in addition to the state-dependent gap $\Delta_{k,q}$ also depends on the eigenvalue $\widetilde{\epsilon}_{k,q}$ of (15) corresponding to the state $k$ and on the chemical potential $\mu_q$, which is determined by the particle number condition given by

$$N_q = \sum_k \widetilde{n}_{k,q}^2, \tag{19}$$

where the occupation number $\widetilde{n}_{k,q}^2$ of the state $k$ is given by,

$$\widetilde{n}_{k,q}^2 = \frac{1}{2}\left[1 - \frac{\widetilde{\epsilon}_{k,q} - \mu_q}{E_{k,q}}\right]. \tag{20}$$

For each type of particles the state-dependent gap in a given state $i$ is obtained as the solution of the so-called gap equation

$$\Delta_{i,q} = -\sum_k v_{i\bar{i},k\bar{k}}^{pair} \frac{\Delta_{k,q}}{2E_{q,k}}. \tag{21}$$

Here, the single particle indices denote the usual quantum numbers, $i \equiv nlj$ and $k \equiv n'l'j'$ for each type of particle. We emphasize that the pairing interaction $v^{pair}$ used in (21) is also determined from the same finite range Gogny interaction (11). The sums over $k$ in Equations (17), (19) and (21) run over bound and quasi-bound states. These quasi-bound states of positive energy are retained by the centrifugal (neutrons) or centrifugal plus Coulomb (protons) barriers [77].

Finally, the total binding energy of a nucleus is given by the sum of the smooth part of the energy computed at VWK level (12) plus the quantal shell correction (16) and the pairing energy (17) calculated perturbatively, i.e.,

$$E_B = E_{\text{VWK}} + \sum_q \left[E_q^{shell} + E_q^{pair}\right]. \tag{22}$$

This method of obtaining the binding energy, which we call VWKSP, was applied for ∼160 even-even nuclei across the whole nuclear chart using three different Gogny forces of D1 type, including D1M* [53]. For D1M*, the relative deviation from the experimental values or the ones obtained with HFB method were found to be within 1%, with only a few exceptions.

*3.3. Outer Crust*

As we have mentioned before, the external region of the NS crust consists of a lattice of fully ionized atomic nuclei embedded in a free electron gas. In the outer layers of the outer crust, the nuclei are the ones which are also observed in terrestrial experiments. However, near the inner crust neutron-rich nuclei whose masses have not been measured experimentally start to appear. To determine the composition and EoS of the outer crust, the essential ingredient is the mass table, which is provided by the experimental masses, when they are known, supplemented by the predictions from theoretical models for the unknown masses. In the present calculation of the outer crust we use the experimental masses from the AME2016 atomic mass evaluation [78] and the recently measured masses of the $^{75-79}$Cu isotopes [79]. When the relevant masses are unknown experimentally, we compute them at HFB level [9] using the D1M and D1M* Gogny interactions. D1M was also used in the calculations of the outer crust of Ref. [80], together with the experimental masses known at that moment (our results with D1M may differ a little from those of Ref. [80] for the layers of the outer crust where new experimental masses available in [78,79] were unmeasured when [80] was published).

The energy of the outer crust at a given density $\rho_{av}$ is computed within the WS approximation, where the energy of each cell containing a nucleus with $Z$ protons and $A$ nucleons has primarily three contributions [81]

$$E(A, Z, \rho_{av}) = E_{Nuc} + E_e + E_{lat}, \quad (23)$$

where, $E_{Nuc}, E_e$ and $E_{lat}$ are the nuclear, electronic and lattice contribution to the energy, respectively. The number density of the outer crust $\rho_{av}$ is determined by the volume $V$ of the cell as $\rho_{av} = A/V$. The nuclear contribution essentially comes from the mass as

$$E_{Nuc} = M(A, Z) = (A - Z)m_n + Zm_p - E_B(A, Z). \quad (24)$$

Here, $m_n$ and $m_p$ are the rest masses of the neutron and the proton, respectively. For masses of nuclei which are not measured experimentally, we use the HFB predictions [9] computed with the D1M* interaction. The electronic contribution $E_e$ is determined by the electronic energy density $\mathcal{H}_e$ for a degenerate relativistic free Fermi gas as

$$E_e = \mathcal{H}_e V, \quad (25)$$

where

$$\mathcal{H}_e = \frac{k_{F_e}}{8\pi^2}\left(2k_{F_e}^2 + m_e^2\right)\sqrt{k_{F_e}^2 + m_e^2} - \frac{m_e^4}{8\pi^2}\ln\left[\frac{k_{F_e} + \sqrt{k_{F_e}^2 + m_e^2}}{m_e}\right], \quad (26)$$

with $m_e$ as the rest mass of electron and $k_{F_e}$ the electron Fermi momentum, which is given by $k_{F_e} = (3\pi^2 n_e)^{1/3}$. In (26) $n_e = (Z/A)\rho_{av}$ is the electron number density. The lattice contribution to the energy is given by

$$E_{lat} = -C\frac{Z^2}{A^{1/3}}k_{F_{av}}, \quad (27)$$

where $k_{F_{av}} = (3\pi^2\rho_{av})^{1/3}$ is the average Fermi momentum connected with the electron Fermi momentum as $k_{F_{av}} = (A/Z)^{1/3}k_{F_e}$ due to charge equilibrium. The constant $C = 0.00340665$ for the bcc lattice is taken from Ref. [82].

At zero temperature, the pressure exerted by the outer crust comes completely from the electrons and the lattice while the nuclei produce no pressure. Therefore,

$$P = -\left(\frac{\partial E}{\partial V}\right)_{A,Z} = P_e + P_{lat} = n_e\sqrt{k_{F_e}^2 + m_e^2} - \mathcal{H}_e - \frac{\rho_{av}}{3}C\frac{Z^2}{A^{4/3}}k_{F_{av}}. \quad (28)$$

To obtain the optimal configuration in a WS cell, we proceed as follows. For a given pressure, at zero temperature, the Gibbs free energy $G$ per nucleon is minimized for different nuclei in the nuclear chart,

$$\begin{aligned} g = \frac{G}{A} &= \frac{E(A,Z,\rho_{av})}{A} + \frac{P}{\rho_{av}} \\ &= \frac{M(A,Z)}{A} + \frac{Z}{A}\sqrt{k_{F_e}^2 + m_e^2} - \frac{4}{3}C\frac{Z^2}{A^{4/3}}k_{F_{av}}. \end{aligned} \quad (29)$$

It is worth mentioning here that recently a new analytical method to evaluate the internal composition of the outer crust has been presented in Ref. [83].

In Figure 6 we plot the composition of the outer crust in terms of the proton number $Z$ and the neutron number $N$ at different average densities $\rho_{av}$, obtained with the nuclear masses measured experimentally (from AME2016 [78] and from [79] for $^{75-79}$Cu) assisted by theoretical HFB calculations [9] using the D1M and D1M* interactions, where the experimental values are not available. For comparison, we also display in the Figure the composition predicted by the BCPM energy density functional [21]. At very low densities (up to $\rho_{av} \sim 10^{-6}$ fm$^{-3}$), the primary contribution comes from Ni and Fe isotopes with neutron numbers $N = 30$, 34 and 36. After that the contribution comes from Kr, Se, Ge and Zn isotopes up to $\rho_{av} \sim 5 \times 10^{-5}$ fm$^{-3}$, with $N = 50$. All three interactions in Figure 6 have the same predictions up to this point because the information primarily comes from the experimental masses. The differences start to appear beyond this density. The elements beyond $\rho_{av} \sim 5 \times 10^{-5}$ fm$^{-3}$ are primarily Ru, Mo, Zr, Sr, Kr or Se isotopes. At these higher densities relevant for the outer crust, the optimal configuration of the WS cell comes from $N = 82$. In the region of the outer crust where the nuclear masses are unknown, the D1M* force predicts the nuclei $^{78}$Ni, $^{128}$Ru, $^{122}$Zr and $^{120}$Se, while the calculations performed with the Skyrme interactions BSk19-BSk21 in Ref. [80] and BSk22 and BSk24-BSk26 in Ref. [23] show a somewhat richer composition, as can be seen in Tables I-III of Ref. [80] and Tables 3–6 of Ref. [23], respectively. The composition of the outer crust critically depends on the nuclear masses, which can be slightly different when computed with different models and extrapolated to the region of unknown masses.

*3.4. Inner Crust*

We resort to the spherical WS approximation for describing the inner crust of NSs. We consider a density range between 0.0004 fm$^{-3}$ and 0.08 fm$^{-3}$ for the inner crust. For the present calculation, we have not considered pasta structures such as cylindrical rods, planar slabs, cylindrical tubes or spherical bubbles, which might be present in between the inner crust and the core of the star. These non-spherical structures may modify the optimal composition of the bottom layers of the inner crust but they do not change the core–crust transition density nor the EoS of the crust in a significant way (see [21] for details). At a given average density of the inner crust, we look for the optimal values of $N$ and $Z$ that satisfy the $\beta$-equilibrium condition

$$\mu_n = \mu_p + \mu_e, \quad (30)$$

where $\mu$ designates the chemical potential of the corresponding particles in the subindex. Once $N$ and $Z$ are fixed, the size of the WS box is determined. The electrons are treated as a free relativistic Fermi gas, with a constant density throughout the WS box. In practice, we proceed as follows. First, we fix the average density and an integer proton number $Z$ and vary the neutron number $N$, which in general is not integer, until the $\beta$-equilibrium condition (30) is reached. Next, keeping the average density fixed, we repeated the procedure for a wide range of $Z$ values searching for the optimal configuration, which corresponds to the WS cell of minimal energy.

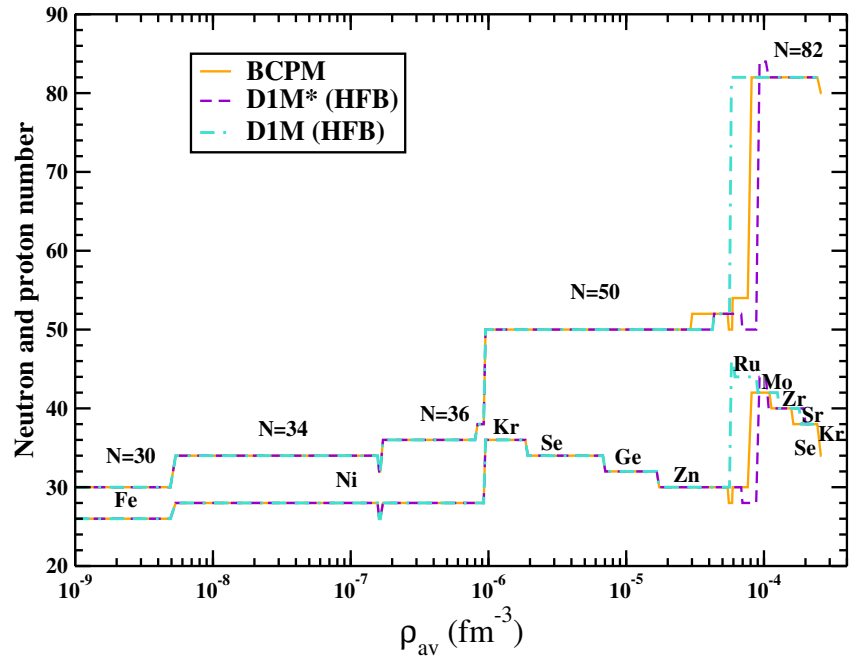

**Figure 6.** Neutron numbers *N* and proton numbers *Z* for the outer crust of NSs with the experimental masses from the AME2016 [78] tabulation plus the recently measured masses of $^{75-79}$Cu [79] aided by theoretical HFB calculations when experimental values are not available, using the D1M and D1M* Gogny forces and the BCPM energy density functional.

For a given *N* and *Z*, we calculate the energy of the WS box with the VWKSP method as explained before for finite nuclei in Section 3.1. We have taken a different form of the density profile for the inner crust unlike the finite nuclei, adapted from Refs. [23,76] as

$$\rho_q(r) = \rho_{B,q} + \frac{\rho_{0,q}}{1 + \exp\left\{\left(\frac{C_q - R_{WS}}{r - R_{WS}}\right)^2 - 1\right\}\exp\left(\frac{r - C_q}{a_q}\right)} \quad . \tag{31}$$

The first term in the right hand side is well suited to obtain a background density at certain average densities of the inner crust. The first exponential in the denominator of the second term is a damping factor tuned by the size of the WS cell ($R_{WS}$), which makes sure that the density reaches the background value (or zero) at the edge of the box. It is worthwhile mentioning here that we added the quantum shell and pairing energies only for protons by the reasons pointed out in [84]. A systematic comparison between the predictions of the extended TF plus Strutinsky integral method including pairing correlations and the fully quantal HFB results demonstrates that the perturbative treatment of shell effects and pairing correlations on top of a self-consistent semiclassical calculation provides a very accurate description of the structure of the NS inner crust [85].

In Figure 7 we plot the binding energy per nucleon (*E*/*A*) subtracted by the bare nucleon mass $m_N$ for 13 different average densities $\rho_{av}$ in the inner crust, which are indicated in the different panels. For comparison, we provide for each average density $\rho_{av}$ the energy obtained in each of the four steps of the calculation of the energy in a WS cell of the inner crust. The orange line with circles denotes the energy containing only the TF contribution, the blue line with squares additionally contains the $\hbar^2$ contributions. The green line with triangles and the red line with diamonds successively take into account the contribution from the shell correction and the pairing energy, respectively. One can clearly observe that once the shell correction is added, the evolution of $E/A - m_N$ produces some local minima. Further addition of the pairing energy (red) somewhat smoothens this feature out. For all average densities but $\rho_{av} = 0.0789$ fm$^{-3}$ the global minimum appears at $Z = 40$. At $\rho_{av} = 0.0789$ fm$^{-3}$ it shifts to $Z = 92$. At $\rho_{av} = 0.0004$ fm$^{-3}$ one can observe shell closures at $Z = 20, 28, 40, 50$, etc., which are similar to ones found in finite nuclei. With the increase in the average density some of these shell closures like $Z = 28$ and 50 are washed away (see the panel with $\rho_{av} = 0.07$ fm$^{-3}$). A systematic study of the inner crust composition performed using the extended TF approach including pairing correlations with a large set of Skyrme forces has been very recently reported [86]. It is shown that the proton content of the WS cells is correlated to the soft or stiff character of the slope of the pure neutron matter EoS for low average densities below 0.05 fm$^{-3}$. In this region the D1M and D1M* interactions predict a relatively stiff neutron matter EoS, which favors Z=40 in the minimal energy configuration (see Figure 7 and Table II of [53]) in agreement with the conclusions drawn in [86].

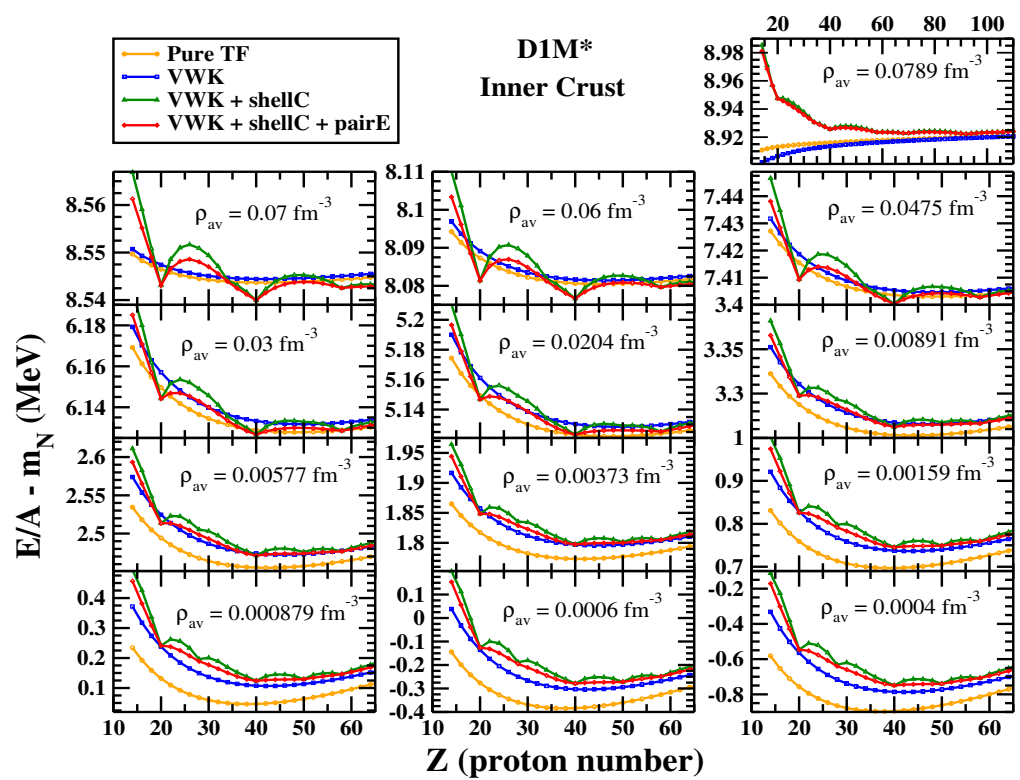

**Figure 7.** Binding energy per nucleon excluding the bare nucleon mass as a function of proton numbers at different average densities $\rho_{av}$ of the inner crust calculated with D1M* Gogny interaction.

In Figure 8 we show the neutron (red solid line) and proton (blue dashed line) density profiles inside the WS cell at different $\rho_{av}$ in the inner crust calculated with the D1M* interaction. With the increase in the $\rho_{av}$, the size of the WS cell shrinks significantly and the cells contain more dense neutron gas. With an increase in the density the diffuseness, particularly for protons, increases significantly. However, the central proton density of the cells increase with decrease in $\rho_{av}$.

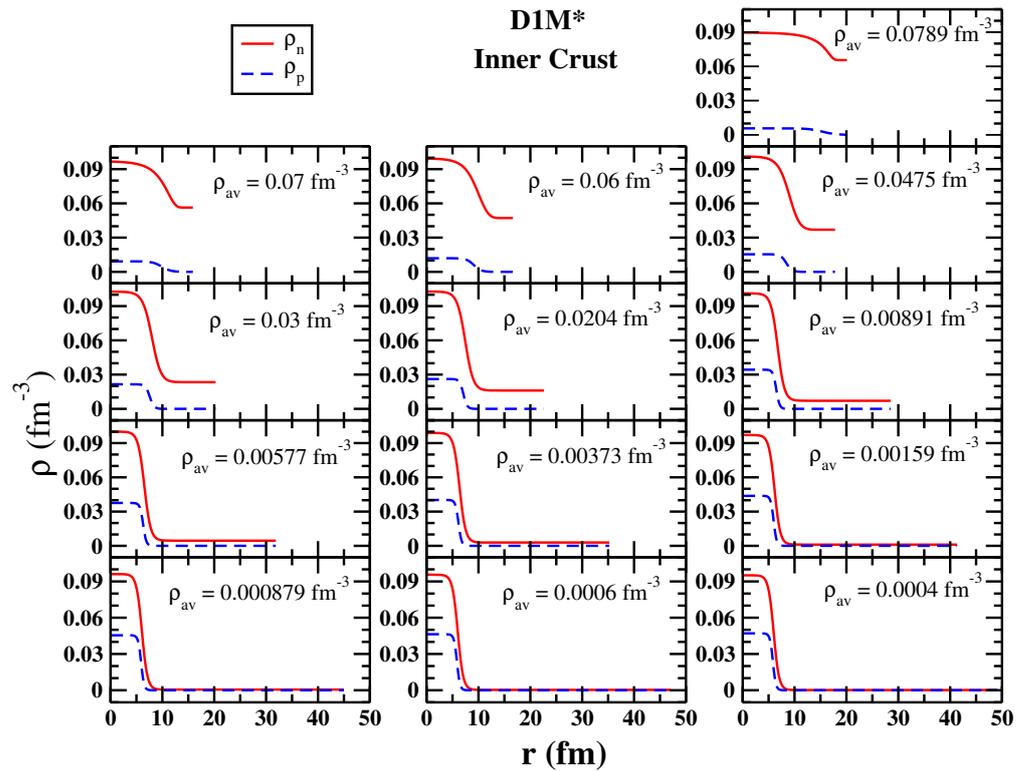

**Figure 8.** Neutron and proton density distribution inside the Wigner–Seitz cells obtained with variational Wigner–Kirkwood method at different average densities $\rho_{av}$ obtained with D1M* Gogny interaction.

*3.5. Core–Crust Transition*

From our calculation in the inner crust we observe that the transition from the crust to the core takes place at an average density around $\sim$0.08 fm$^{-3}$. To find the core–crust transition density within a given model requires, in principle, the computation of the complete EoS of the inner crust, which is not a simple task, as we have seen along this section. However, the search of the crust–core transition density can be considerably simplified by performing the calculation from the core side. In this case one searches for the violation of the stability conditions of the homogeneous core under small amplitude oscillations, which indicate the appearance of nuclear clusters and therefore the transition to the inner crust. There are different ways to determine the core–crust transition from the core side, namely the thermodynamical method ($V_{ther}$), the dynamical method ($V_{dyn}$), random phase approximation and the Vlasov equation method (see Ref. [13,20,87] for more details and further references).

In the thermodynamical method the stability of the NS core is discussed in terms of bulk properties only, where the mechanical and chemical stability conditions set the boundaries of the homogeneous core:

$$-\left(\frac{\partial P}{\partial v}\right)_{\mu_{np}} > 0, \quad -\left(\frac{\partial \mu_{np}}{\partial q}\right)_{v} > 0, \tag{32}$$

where $P$ is the total pressure of neutron star matter (1)–(2), $\mu_{np}$ is the difference between the neutron and proton chemical potentials, $v = 1/\rho$ is the inverse of the baryon density and $q$ is the charge per baryon. In the low density regime of interest for the core–crust transition the chemical stability is always fulfilled and the mechanical stability condition can be recast through the so-called thermodynamical potential $V_{ther}(\rho)$ [13,20]. The thermodynamical potential is a function of the baryon density only and the transition density corresponds to the value of $\rho$ for which $V_{ther}(\rho)$ changes sign (see [13] and references therein).

The dynamical method, introduced in Ref. [81], assumes that the nuclear energy density can be expressed as the sum of a bulk homogeneous part and an inhomogeneous contribution, which depends on the gradient of the neutron and proton densities as well as on the direct part of the Coulomb potential. The Skyrme forces fit this scheme [88]. However, for finite-range interactions, such as the Gogny forces, the calculation is more involved. Quite often the energy density functional for finite-range forces can be approximated very accurately by a local form using the extended TF density matrix [72] instead of the full HF density matrix. Within this scheme, the energy density can be written as a homogeneous term, provided by the Slater density matrix ($\hbar^0$ term), plus an additional $\hbar^2$ contribution written in terms of the gradients of the neutron and proton densities and of the inverse of the momentum and position dependent effective masses [87]. This inhomogeneous contribution also contains the gradient expansion of the direct nuclear and Coulomb potentials (see [20,87] for more details). Thus,

$$E = E_0 + \frac{1}{2} \sum_{i,j} \int \frac{d\mathbf{k}}{(2\pi)^3} \frac{\delta^2 E}{\delta n_i(\mathbf{k}) \delta n_j^*(\mathbf{k})} \delta n_i(\mathbf{k}) \delta n_j^*(\mathbf{k}), \tag{33}$$

where $E_0$ is the unperturbed density and $n_i(\mathbf{k})$ are the momentum distributions (inverse Fourier transform of the density perturbation) for each type of particles. The second variation of the energy defines the so-called curvature matrix, which is the sum of three different terms. The first is the bulk contribution, which defines the stability of uniform NS matter and corresponds to the equilibrium condition in the thermodynamical method. The second term collects the gradient contributions in the energy density functional and is a function of the momentum *k*. For zero-range Skyrme forces it is a quadratic function [88], but it is a more involved function in the case of finite-range interactions [20,87]. The last contribution is due to the direct Coulomb interactions between protons and electrons. The stability condition requires the curvature matrix to be convex. This allows one to write a dynamical potential $V_{dyn}(\rho, k)$, which is now momentum- and density-dependent. To compute the transition density one first minimizes for each value of the density $\rho$ the dynamical potential respect to $k$. Next, as in the case of the thermodynamical method, one determines the transition density as the value of the density for which $V_{dyn}(\rho, k(\rho))$ vanishes (see Refs. [20,87] for a detailed description of the dynamical method). Table 3 collects the main core–crust transition properties, namely density, pressure and isospin asymmetry, derived with the thermodynamical and dynamical methods using the D1M, D1M* and D1M** Gogny forces as well as with the BCPM energy density functional, which is used here as a benchmark.

It is known from earlier literature that the core–crust transition density, estimated in the thermodynamical approach, using Skyrme and Relativistic Mean Field (RMF) models, shows a decreasing trend with an increasing value of the slope of the symmetry energy (see [13,87] and references therein). In Refs. [13,87] we have computed the core–crust transition density predicted by finite-range interactions using the thermodynamical and dynamical methods. We find that our results are in harmony with earlier calculations obtained with the Skyrme interactions and RMF parametrizations. This can be seen in Figure 9, where we plot the transition density (left panels) and the transition pressure (right panels) obtained using the thermodynamical (upper panels) and the dynamical (lower panels) methods. We have obtained the transition properties for a large set of Skyrme forces and also for most of Gogny interactions available in the literature. These sets of interactions cover a large range of values of the slope of the symmetry energy *L* going from around 15 MeV up to 130 MeV. We see that the values of both the transition density and the transition pressure have larger values when they are obtained using the thermodynamical method instead of the dynamical method. The reason behind this is, as we have mentioned, that the dynamical method takes into account the surface and Coulomb contributions that tend to stabilize more the liquid core. Comparing between the transition density and pressure we observe different behaviors. On the one hand the values of the density of the core–crust transition follow a rather linear decreasing trend with respect to the slope of the symmetry energy *L* for both Skyrme and Gogny forces. On the other hand, the correlation between the transition pressure and *L* is less obvious, being more visible for Skyrme forces than for the Gogny ones. For example, we can see from Table 3 the decreasing trend of the transition density with the increasing value of *L* of the different models considered in this Table (see Table 2 in this respect), while the transition pressure is roughly similar computed with the D1M, D1M* and D1M** forces and differs from the prediction of the BCPM energy density functional.

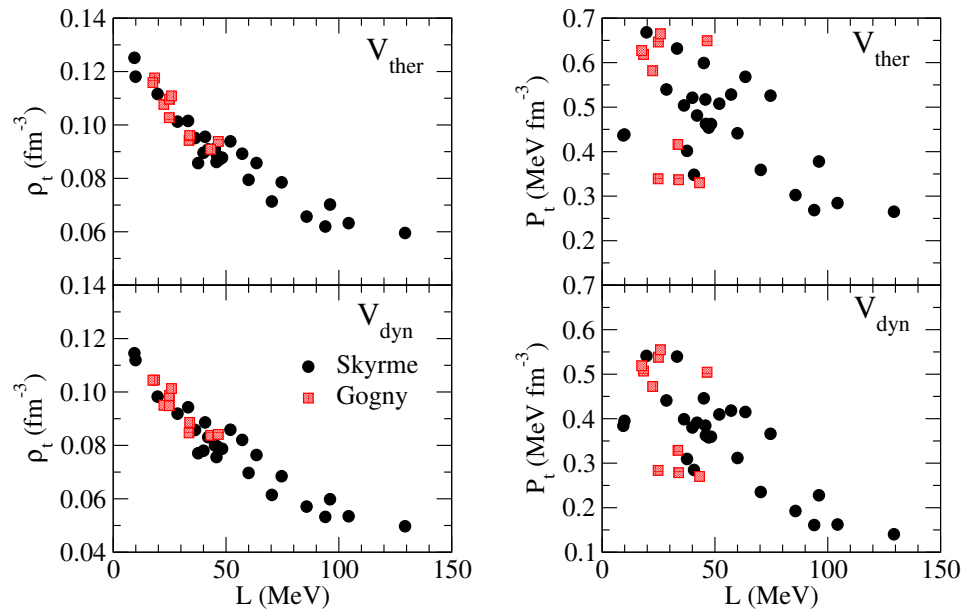

**Figure 9.** Transition density (left panels) and transition pressure (right panels) against the slope of the symmetry energy computed for some Skyrme and Gogny interactions. The upper panels correspond to the values obtained using the thermodynamical method whereas the lower panels display the results extracted using the dynamical method.

**Table 3.** Core–crust transition density $\rho_t$, pressure $P_t$ and and isospin asymmetry $\delta_t$ predicted by the D1M, D1M* and D1M** Gogny forces and the BCPM energy density functional.

|  | $\rho_t$ (fm$^{-3}$) | $P_t$ (MeVfm$^{-3}$) | $\delta_t$ |
|---|---|---|---|
| D1M |  |  |  |
| $V_{ther}$ | 0.1027 | 0.3390 | 0.9241 |
| $V_{dyn}$ | 0.0949 | 0.2839 | 0.9257 |
| D1M* |  |  |  |
| $V_{ther}$ | 0.0909 | 0.3301 | 0.9275 |
| $V_{dyn}$ | 0.0838 | 0.2702 | 0.9300 |
| D1M** |  |  |  |
| $V_{ther}$ | 0.0960 | 0.3368 | 0.9257 |
| $V_{dyn}$ | 0.0886 | 0.2786 | 0.9279 |
| BCPM |  |  |  |
| $V_{ther}$ | 0.0889 | 0.5137 | 0.9339 |
| $V_{dyn}$ | 0.0816 | 0.4132 | 0.9382 |

## 4. Global Properties of Neutron Stars Predicted by Gogny Forces

The unified EoS is obtained from the consistent calculation of the core and the crust, as we have shown in the previous sections. We provide the unified EoSs and the associated stellar matter composition obtained for D1M and D1M* in the supplementary material. In addition, for the sake of clarity we display in the left panel of Figure 10 the unified EoS in logarithmic scale computed with these interactions as well as with the BCPM energy density functional. From this panel we see that practically no differences can be observed in the outer crust. In the inner crust the EoSs provided by the different Gogny forces are similar and show some differences with the BCPM predictions. However, in the core region the differences between the original D1M and modified D1M* Gogny forces are more prominent.

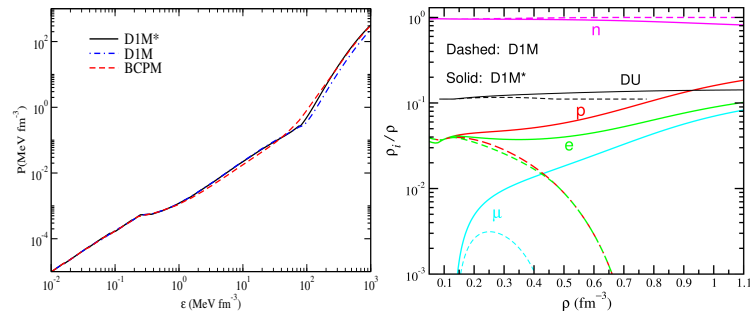

**Figure 10.** Left: Unified EoS computed with the the D1M and D1M* Gogny force and with the BCPM energy density functional. Right: Particle fractions and the proton fraction corresponding to the onset of the direct Urca (DU) process (see text for details) as functions of the nucleonic density from the D1M and D1M* interactions.

In the right panel of Figure 10 we compare the predictions of the D1M and D1M* EoSs for the particle populations in the beta-equilibrated $npe\mu$ matter of the NS core. The impact of the stiffer symmetry energy of the D1M* interaction with respect to D1M, as reflected by the total EoS displayed in the left panel of the Figure, can be clearly seen in the right panel. D1M* predicts a persistent population of protons and leptons in the core of the star with increasing nucleon density. In stark contrast, in the results calculated with D1M we see that matter becomes soon deprotonized and deleptonized when the density increases. This is because in D1M it is much less costly to convert protons into neutrons due to the softer symmetry energy of this interaction. Actually, according to D1M the stellar core would be composed practically of only neutrons after a density $\rho \approx 0.65$ fm$^{-3}$ ($\approx 4\rho_0$), as can be seen from the D1M particle fractions in Figure 10. Notice also that D1M* predicts a growing population of muons with higher density, whereas in D1M the appearance of muons is nominal. According to recent studies in the literature, the presence of muons in NSs may play a significant role in addressing several new physics questions about the interactions and the astrophysical effects of muonphilic dark matter particles, see Ref. [89] and references therein. The proton fraction inside the beta-equilibrated matter also determines whether a proto-neutron star will go through the direct Urca process or not. In $npe\mu$ matter this is attributed to the condition that the proton fraction satisfies $\rho_p/\rho > x_{DU}$, where $x_{DU}$ is defined as [90]

$$x_{DU} = \left[ 1 + \left\{ 1 + \left( \frac{\rho_e}{\rho_p} \right)^{1/3} \right\}^3 \right]^{-1}. \tag{34}$$

In Figure 10, we plotted this quantity as a function of density, denoted by "DU" (black lines). The density point at which the proton fraction (red) surpasses the quantity $x_{DU}$ indicates the onset of direct Urca. One can see that only D1M* fulfills this condition, though at fairly large densities ($\rho > 0.93$ fm$^{-3}$). This behavior can be directly attributed to the stiffer symmetry energy for D1M* at suprasaturation densities compared to D1M.

Once the full EoS is obtained, one can look for different global properties of NSs. In this review we will concentrate on three relevant aspects, namely the mass-radius relation in an NS, which provides a detailed information about the structure of the star, the moment of inertia of the NS, and in particular its fraction enclosed by the crust, which may be important for the description of pulsar glitches. Finally, the last aspect to be discussed is the tidal deformability in binary systems of NS. This quantity can be accessed by the detection of gravitational waves (GW), coming for example from the merger of a NS binary as in the GW170817 event recorded recently.

*4.1. The Tolman–Oppenheimer–Volkov Equations*

In order to study the mass-radius relation of NSs, one has to solve the TOV equations [54,91], which need as an input the full EoS along all of the star. The TOV equations take into account within the general relativity framework the hydrostatic equilibrium in the star between the pressure given by the gravitational field and the pressure coming from the baryons and leptons inside the star. The TOV equations are given by

$$\frac{dP(r)}{dr} = \frac{G}{r^2 c^2}[\epsilon(r) + P(r)]\left[m(r) + 4\pi r^3 P(r)\right]\left[1 - \frac{2Gm(r)}{rc^2}\right]^{-1} \quad (35)$$

$$\frac{dm(r)}{dr} = 4\pi r^2 \epsilon(r), \quad (36)$$

where $\epsilon(r)$, $P(r)$ and $m(r)$ are, respectively, the energy density (including free nucleon mass), pressure and mass at each radius $r$ inside the NS. Starting with a central energy density $\epsilon(0)$, a central pressure $P(0)$ and a central mass $m(0) = 0$, one integrates outwards the differential equations until reaching the NS surface, where the pressure is zero, $P(R) = 0$. At the same time, the location of the surface of the star determines its total radius $R$ and its total mass $M = m(R)$.

In Figure 11 we plot the mass–radius (MR) relation for the D1M and D1M* Gogny interactions, as well as for the BCPM energy density functional. We stress that all three EoSs used in the calculations are unified EoSs, where the outer crust, the inner crust and the core have been obtained using the same interaction. In the same plot we include constraints coming from different sources. First, we include constraints for the maximum mass obtained from the observation of the highly massive NSs [14,15]. The green vertical constraint comes from cooling tails of type-I X-ray bursts in three low-mass X-ray binaries and a Bayesian analysis [92], and the blue vertical constraint is from five quiescent low-mass X-ray binaries and five photospheric radius expansion X-ray bursters after a Bayesian analysis [93]. The pink–red rectangular constraint at the front is from a Bayesian analysis with the data from the GW170817 detection of gravitational waves from a binary NS merger [94]. Finally, we inserted the constraints coming from the very recent NICER observations for the mass and radius of the pulsars PSR J0030+0451 and PSR J0740+6620 with one-sigma deviation [95,96]. As mentioned in previous sections, we observe that the D1M interaction predicts the NS maximum mass of only $1.74 M_\odot$. Moreover, the MR relation obtained from D1M falls outside all considered constraints. If we look at the MR relation obtained using the EoS given by the D1M* interaction, we see that it reaches a maximum NS mass of around $2M_\odot$, similarly to the one given by the BCPM energy density functional, which we included here as a benchmark. The MR relations given by both D1M* and BCPM lie inside most of the constraints for the mass and radius included in the same Figure.

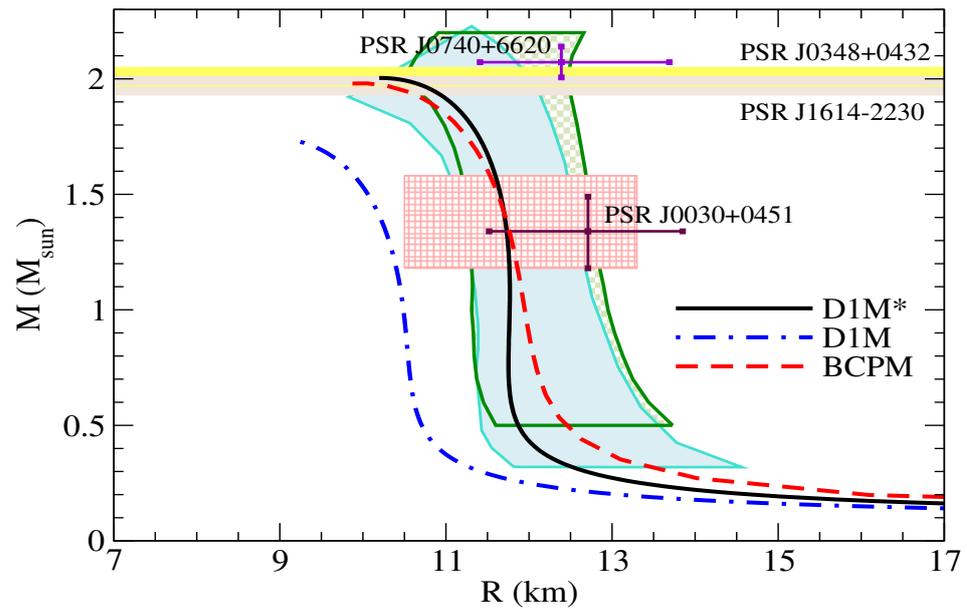

**Figure 11.** Mass-radius relation obtained using the D1M* and the D1M Gogny forces and the BCPM energy density functional. Constraints from the measurements of $M \approx 2 M_\odot$ (yellow and grey) [14,15], from cooling tails of type-I X-ray bursts in three low-mass X-ray binaries and a Bayesian analysis (green) [92], from five quiescent low-mass X-ray binaries and five photospheric radius expansion X-ray bursters after a Bayesian analysis (blue) [93] and from a Bayesian analysis with the data from the GW170817 detection of gravitational waves from a binary NS merger (red) [94] are shown. Finally, the very recent constraints coming from the NICER mission are also included [95,96].

We plot in Figure 12 the mass (left panel) and radius (central panel) enclosed in the NS crust. The values of the crustal mass for the BCPM energy density functional are larger than the ones obtained using Gogny interactions, but are close to the ones computed with D1M* once one approaches the NS maximum mass values. On the other hand, the crustal masses obtained using the D1M interaction are lower than the ones obtained with D1M*. For the crustal radius, or thickness of the crust, we see that the values predicted by D1M* are very similar to the results one achieves for BCPM, at least above $1.4 M_\odot$, while the crustal radius computed with D1M is smaller than that for the D1M* interaction or for the BCPM energy density functional.

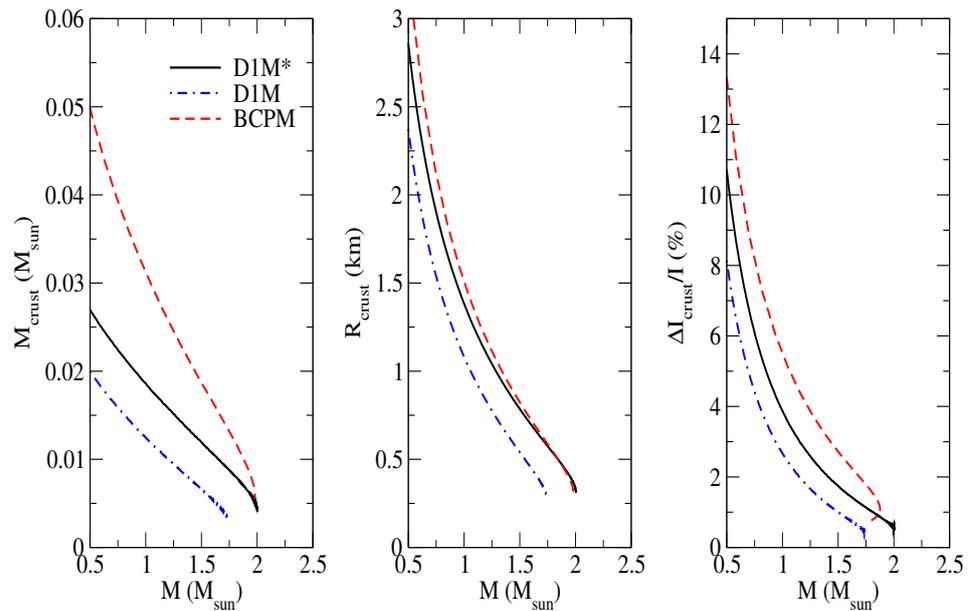

**Figure 12.** Crustal mass (**left**), crustal radius (**center**), and crustal fraction of the moment of inertia ($\Delta I_{\rm crust}/I$) (**right**) obtained with the D1M*, D1M and BCPM interactions.

*4.2. Moment of Inertia*

The moment of inertia of slowly-rotating NSs can be computed from the static mass distribution and the gravitational potentials that one finds when solving the TOV equations [97]. If one studies the slow-rotation limit, the moment of inertia is given by [81,97,98]

$$I \equiv \frac{J}{\Omega} = \frac{8\pi}{3}\int_0^R r^4 e^{-\nu(r)}\frac{\bar{\omega}(r)}{\Omega}\frac{(\epsilon(r)+P(r))}{\sqrt{1-2Gm(r)/rc^2}}dr, \qquad (37)$$

where $G$ is the gravitational constant and $c$ the speed of light and one has assumed spherical symmetry. In Equation (37), $J$ is the angular momentum, $\Omega$ is the stellar rotational frequency, $\nu(r)$ and $\bar{\omega}$ are radially dependent metric functions and $m(r)$, $\epsilon(r)$ and $P(r)$ are, respectively, the NS mass, energy density and total pressure enclosed in a radius $r$. The metric function $\nu(r)$ satisfies [98]

$$\nu(r) = \frac{1}{2}\ln\left(1-\frac{2GM}{Rc^2}\right) - \frac{G}{c^2}\int_r^R \frac{(M(x)+4\pi x^3 P(x))}{x^2(1-2GM(x)/xc^2)}dx, \qquad (38)$$

and the angular velocity of the fluid measured in a local reference frame is given by the relative frequency $\bar{\omega}(r) \equiv \Omega - \omega(r)$, where $\omega(r)$ is the frequency that appears because of the slow rotation of the star. On the other hand, the relative frequency $\tilde{\omega}(r) \equiv \bar{\omega}(r)/\Omega$ can be obtained by solving the differential equation [98]

$$\frac{d}{dr}\left(r^4 j(r)\frac{d\tilde{\omega}(r)}{dr}\right) + 4r^3\frac{dj(r)}{dr}\tilde{\omega}(r) = 0, \qquad (39)$$

with

$$j(r) = \begin{cases} e^{\nu(r)}\sqrt{1-2Gm(r)/rc^2} & \text{if } r \leq R \\ 1 & \text{if } r > R \end{cases}. \qquad (40)$$

The relative frequency $\tilde{\omega}(r)$ obtained as a solution of (39) and (40) has to fulfill the following boundary conditions

$$\tilde{\omega}'(0) = 0 \qquad \text{and} \qquad \tilde{\omega}(R) + \frac{R}{3}\tilde{\omega}'(R) = 1. \qquad (41)$$

Notice that in the slow-rotation regime the solution of the moment of inertia does not depend on the stellar frequency $\Omega$. Starting from an arbitrary value of $\tilde{\omega}(0)$, one integrates Equation (39) up to the surface. Usually, it will be necessary to re-scale the function $\tilde{\omega}(r)$ and its derivative with an appropriate constant in order to fulfill (41). One can test the accuracy of the final result by checking the condition [98]

$$\tilde{\omega}'(R) = \frac{6GI}{R^4 c^2}. \tag{42}$$

The ratio between the fraction of the moment of inertia $\Delta I_{\text{crust}}$ and the total moment of inertia $I$ is intrinsically connected to pulsar glitches and to the location of the core–crust transition [31,54,99,100]. We plot in the right panel of Figure 12 the ratio between $\Delta I_{\text{crust}}/I$ against the total NS mass for the D1M and D1M$^*$ interactions and the BCPM energy density functional. Similarly to what happens for the crustal mass and crustal radius, the crustal fraction of the moment of inertia is larger when obtained using the BCPM EoS. On the other hand, the values that one obtains with D1M$^*$ fall between the ones of BCPM and the ones given by D1M, which provides the lower values of $\Delta I_{\text{crust}}/I$ from these three interactions. As can be seen in the rightmost panel of Figure 12, the values obtained using the D1M$^*$ interaction lie between the results predicted by the BCPM and D1M EoSs. Notice that this later provides the lower values of the ratio $\Delta I_{\text{crust}}/I$ among all the interactions used in this calculation. To account for the size of the pulsar glitches, the pinning model requires that some amount of angular momentum is carried out by the crust, which can be recast as a constraint on the crustal fraction of the moment of inertia. For example, to explain Vela and another source of glitches, first estimates suggested that $\Delta I_{\text{crust}}/I > 1.4$ % [101], although more recent estimates, which take into account the neutron entrainment in the crust, increases the minimal crustal fraction up to 7% in order to explain the glitching phenomena [102,103]. When the Gogny forces D1M and D1M$^*$ are used to evaluate the moment of inertia, the first constraint is fulfilled for NS with masses smaller than 1.4 and 1.7$M_\odot$, respectively, while the second constraint is only fulfilled by very small NS masses, as can be seen in the rightmost panel of Figure 12. If the calculation of the moment of inertia is performed using the BCPM energy density functional instead of the D1M and D1M$^*$ forces, the behavior is similar, although the glitching sources have slightly larger masses.

The left panel of Figure 13 encloses the total moment of inertia against the total NS mass for the same interactions as the previous Figure. The values of the moment of inertia obtained with D1M$^*$ and BCPM are very similar from low masses up to 1.5$M_\odot$, from where the moment of inertia computed with D1M$^*$ is slightly larger than that for BCPM. For these two interactions, the maximum values of the moment of inertia are $1.95 \times 10^{45}$ g cm$^2$ and $1.88 \times 10^{45}$ g cm$^2$, respectively, which are reached a little bit before the maximum mass configuration. Contrary to these two interactions, the D1M Gogny force gives much smaller values for $I$, reaching maximum values of only $1.30 \times 10^{45}$ g cm$^2$. It is expected that binary pulsar observations can provide new information about the moment of inertia and, therefore, put additional constraints on the EoS of NS [100]. The moment of inertia of the primary component of the pulsar PSR-J0737-3039, which has a mass of 1.338$M_\odot$, has been estimated by Landry and Kumar in the range $I = 1.15^{+0.38}_{-0.24} \times 10^{45}$ g cm$^2$ [104]. From the left panel of Figure 13 it can be seen that this constraint is fulfilled by the moment of inertia computed using the EoSs based on the D1M and D1M$^*$ forces and the BCPM energy density functional (see Ref. [20] for more details). Finally, let us mention that the dimensionless quantity $I/MR^2$ is found to scale with the NS compactness $\chi = GM/Rc^2$ and to be almost independent of the mass and radius of the NS [99,100,105]. We checked that this is the situation when the moment of inertia is computed using the D1M and D1M$^*$ and the BCPM energy density functional on the one hand, and also that the universal relation $I/MR^2$ vs. $\chi$ lies within the region estimated by Lattimer and Schutz [100] and Breu and Rezolla [105] when studied with the same interactions.

*4.3. Tidal Deformability*

The detection of GW coming from mergers of binary NS systems, and of NS–Black Hole systems, will open new possibilities to study the EoS of highly asymmetric nuclear matter, which one uses to describe the interior of NSs. If we focus on binary NS systems, each of its components induces a gravitational tidal field on its companion. This phenomenon leads to a mass-quadrupole deformation on each member of the binary. To linear order, the tidal deformability $\Lambda$ describes this tidal deformation of each star in the system, and it is defined as the ratio between the induced quadrupole moment and the external tidal field [106,107].

For each of the stars in the binary, the tidal deformability is given by [106–108]

$$\Lambda = \frac{2}{3} k_2 \left( \frac{Rc^2}{GM} \right)^5, \tag{43}$$

where $k_2$ is the dimensionless tidal Love number, $R$ is the NS radius, $M$ its total mass. As previously stated in this paper, the solution of the TOV equations provides the values of the mass and radius of a NS, while the Love number $k_2$ is obtained as

$$\begin{aligned}
k_2 &= \frac{8\chi^5}{5}(1-2\chi)^2[2+2\chi(y-1)-y] \times \Big\{ 2\chi[6-3y+3\chi(5y-8)] \\
&+ 4\chi^3\Big[13-11y+\chi(3y-2)+2\chi^2(1+y)\Big] \\
&+ 3(1-2\chi)^2[2-y+2\chi(y-1)]\ln(1-2\chi) \Big\}^{-1},
\end{aligned} \tag{44}$$

where $\chi$ is the compactness of the star and

$$y = \frac{R\beta(R)}{H(R)}. \tag{45}$$

In Equations (44) and (45), $\beta(R)$ and $H(R)$ are given by the solution of the following set of coupled differential equations [107,108]:

$$\frac{dH(r)}{dr} = \beta(r) \tag{46}$$

$$\begin{aligned}
\frac{d\beta(r)}{dr} &= \frac{2G}{c^2}\left(1-\frac{2Gm(r)}{rc^2}\right)^{-1} H(r)\bigg\{-2\pi\left[5\epsilon+9p+\frac{d\epsilon}{dp}(\epsilon+p)\right]+\frac{3c^2}{r^2 G} \\
&+ \frac{2G}{c^2}\left(1-\frac{2Gm(r)}{rc^2}\right)^{-1}\left(\frac{m(r)}{r^2}+4\pi rp\right)^2\bigg\} \\
&+ \frac{2\beta(r)}{r}\left(1-\frac{2Gm(r)}{rc^2}\right)^{-1}\bigg\{-1+\frac{Gm(r)}{rc^2}+\frac{2\pi r^2 G}{c^2}(\epsilon-p)\bigg\},
\end{aligned} \tag{47}$$

where $m(r)$ is the mass enclosed inside a radius $r$, and $\epsilon$ and $p$ are the corresponding energy density and pressure. One solves Equations (46) and (47) along with the TOV equations by integrating outwards, with the boundary conditions $H(r) = a_0 r^2$ and $\beta(r) = 2a_0 r$ as $r \to 0$. The constant $a_0$ is arbitrary, as it cancels out in the expression for the Love number [108].

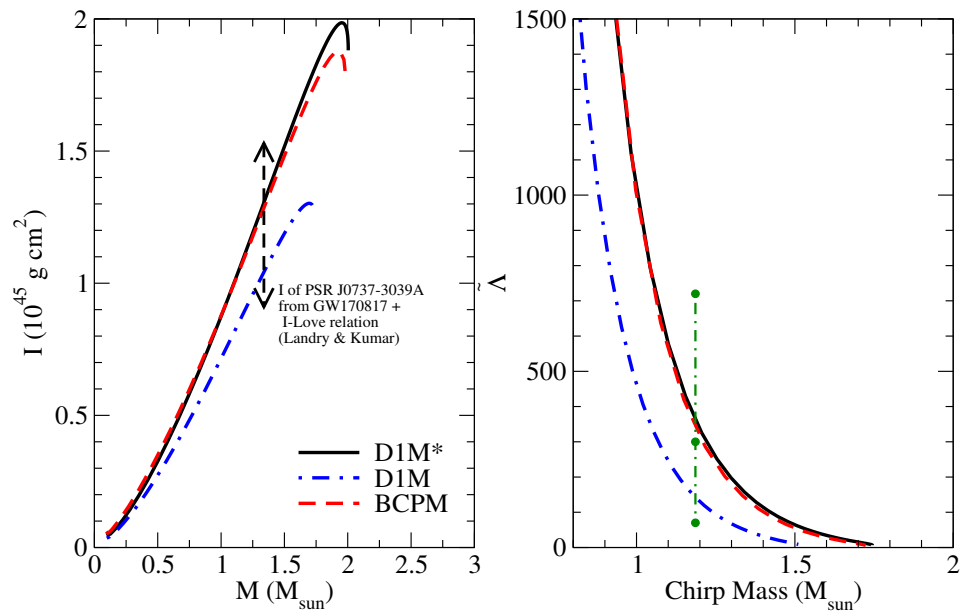

**Figure 13.** Left: Total moment of inertia against the total mass of neutron stars computed using the D1M* and D1M Gogny forces and the BCPM energy density functional. The constraint proposed in [104] is also displayed. Right: Mass weighted tidal deformability (for symmetric binaries) against the chirp mass of binary neutron star systems obtained using the same interactions as in the left panel. The constraint for $\tilde{\Lambda}$ coming from the GW170817 event is also included [109,110].

When studying the full NS binary system, the mass-weighted tidal deformability $\tilde{\Lambda}$ takes into account the contribution from the tidal effects to the phase evolution of the gravitational wave spectrum of the inspiraling NS binary, and it is defined as

$$\tilde{\Lambda} = \frac{16}{13} \frac{(M_1 + 12M_2)M_1^4\Lambda_1 + (M_2 + 12M_1)M_2^4\Lambda_2}{(M_1 + M_2)^5}, \quad (48)$$

where $\Lambda_1$ and $\Lambda_2$ are the tidal deformabilities of each NS conforming the system and $M_1$ and $M_2$ are their corresponding masses. Notice that the definition (48) fulfills $\tilde{\Lambda} = \Lambda_1 = \Lambda_2$ when $M_1 = M_2$.

The LIGO and Virgo Collaboration have already detected some GW signals coming from the merger of two NSs [109,110], which allow constraining of the mass-weighted tidal deformability $\tilde{\Lambda}$ and the chirp mass of the system $\mathcal{M}$, which is given by

$$\mathcal{M} = \frac{(M_1 M_2)^{3/5}}{(M_1 + M_2)^{1/5}}. \quad (49)$$

In this paper we will use the constraints coming from the GW170817 detection [109,110], as it is at the moment the one that further constrains $\tilde{\Lambda}$ and $\mathcal{M}$, at values of $\tilde{\Lambda} = 300^{+420}_{-230}$, $\mathcal{M} = 1.186^{+0.001}_{-0.001} M_\odot$. Additional constraints for the single NS masses are also given as $M_1 \in (1.36, 1.60) M_\odot$ and $M_2 \in (1.16, 1.36) M_\odot$.

We plot in the right panel of Figure 13 the mass-weighted tidal deformability against the chirp mass obtained using the D1M* and D1M Gogny forces and the BCPM energy density functional. The mass-weighted tidal deformability $\tilde{\Lambda}$ predicted by the BCPM and D1M* EoSs have very similar values, lying well inside the constraint of the GW170817 detection, which is plotted in green in the same Figure. On the other hand, the values obtained with the D1M Gogny interaction are lower than the ones obtained with D1M* and BCPM, even though they also lie inside the GW constraints, but near the lower limit. Finally, let us mention that in Ref. [111] an analysis of the GW170817 constraints has been performed using both Gogny forces and momentum-dependent interactions (MDI). One of the conclusions of this study has been that the successful Gogny and MDI interactions that are compatible with GW170817 restrict the radius of a canonical NS of $1.4 M_\odot$ to within the range of 9.4 km $\leq R_{1.4} \leq$ 13.1 km [111].

## 5. Conclusions

In this review article we revised and summarized the most relevant aspects of our investigations about the application of effective forces of Gogny type to the NS scenario that have been previously reported in a series of papers. The Gogny interactions were proposed more than forty years ago with the purpose to describe simultaneously the mean field and the pairing field, which usually are disconnected in almost all of the mean field models available in the literature. Although the standard parametrizations of the Gogny force, such as D1S, D1N and D1M, reproduce rather accurately the nuclear masses as well as pairing and deformation properties of finite nuclei, these interactions fail when applied to the NS domain. The basic reason for that is the too soft symmetry energy predicted by these forces at high baryon densities, which are unable to produce heavy enough stellar masses. To cure this limitation, we proposed a reparametrization of the Gogny D1M force in such a way that preserves the accurate description of finite nuclei, the isovector properties of the interaction, in particular the slope of the symmetry energy are modified to obtain a stiffer EoS able to predict maximal NS masses of about $2M_\odot$, in agreement with well-contrasted astronomical observations. Our renormalization procedure has been applied using the D1M force as starting point, because the D1S and D1N interactions are too far from the $2M_\odot$ target. In this way we have built up two new Gogny parametrizations, denoted as D1M$^*$ and D1M$^{**}$, which predict maximal masses of NS of $2M_\odot$ and $1.91M_\odot$, respectively.

Apart from the description of the core of NSs, we also used these new Gogny forces to build up the EoS of the crust of NSs aimed to obtain a unified EoS from the surface to the center of the star. The outermost region of a NS, called outer crust, consists of a lattice of atomic nuclei, which are more neutron rich as the depth increases, embedded in a free electron gas. The basic ingredient to determine the EoS in this region are the nuclear masses, which are taken from the experiment or obtained from a HFB calculation with the same Gogny force when the masses are unknown. After a density around 0.003 fm$^{-3}$, neutrons cannot be retained by the nuclei and above this density, the matter is arranged still as a lattice structure but now permeated by free neutron and electron gases. The treatment of this region is complicated owing to the presence of the neutron gas. To describe this scenario, called inner crust, we use the Wigner–Seitz approximation and compute the representative nucleus inside each cell using the semiclassical Variational Wigner–Kirkwood approximation, which includes $\hbar^2$ corrections added perturbatively. Moreover, the quantal shell corrections and the pairing correlations for protons are also added perturbatively, using the so-called Strutinsky integral method and the BCS approximation, respectively. At a density roughly around one-half the saturation density the inner crust structure dissolves in a homogeneous core. The precise value of the crust–core transition density is strongly model dependent. To determine the transition point is not an easy task when looking from the crust, as it requires an accurate description of the inner crust. However, it is easier to determine the transition point from the core side searching for the density for which the homogeneous core is unstable against the cluster formation. The simplest approach is the so-called thermodynamical method that only considers the stability of the homogeneous core. A more precise approximation is provided by the dynamical method, which on top of the stability of the homogeneous core, also considers finite-size effects. We have shown that the dynamical method predicts transition densities and pressures in rather good agreement with the estimate obtained from the crust side.

Once the full EoS based on the modified D1M$^*$ Gogny force was obtained, we used it to predict different NS properties. In addition to the mass–radius relation, we analyzed the behavior of the moment of inertia and the tidal deformability of the star, which can be related to information extracted from observations in some binary pulsars and from the GW170817 event. We also analyzed some global crustal properties such as the mass and radius of the crust, as well as the crustal fraction of the moment of inertia, which can be relevant for the description of the glitches. We find that these global properties obtained with the Gogny-based EoS are in good agreement with the predictions of other well contrasted EoS as the ones based on the SLy4 Skyrme force [22] or the microscopic BCPM energy density functional [21], which is used as a benchmark in this work. Although a detailed study of some other nuclear structure phenomena, such as the description of odd nuclei, fission phenomena or giant resonances computed with the new D1M$^*$ and D1M$^{**}$ Gogny forces is still pending, we conclude that these new interactions are promising alternatives to describe simultaneously finite nuclei and neutron stars providing results in harmony with the experimental data and astronomical observations.

**Author Contributions:** All the authors contributed equally to the writing of the review with X. Viñas leading the coordination of all different tasks. All authors have read and agreed to the published version of the manuscript.

**Funding:** C.G., M.C. and X.V. were partially supported by Grants No. FIS2017-87534-P, No. PID2020-118758GB-I00 (DOI 10.13039/501100011033) and No. CEX2019-000918-M (through the "Unit of Excellence María de Maeztu 2020-2023" award to ICCUB) from the State Agency for Research (AEI) of the Spanish Ministry of Science and Innovation (MICINN). The work of C.M. was partially supported by the IN2P3 Master Project "NewMAC". The work of L.M.R. was partly supported by the Spanish MINECO Grant No. PGC2018-094583-B-I00.

**Institutional Review Board Statement:** This manuscript has gone through VIRGO document review system as C.M. is obliged to make sure any infringement of the VIRGO data has not taken place, being part of the VIRGO collaboration through Caen-Meudon group.

**Informed Consent Statement:** Not applicable.

**Data Availability Statement:** See supplemental material accompanying this paper, further inquiries can be directed to the corresponding author.

**Acknowledgments:** The authors are indebted to J.N. De for a careful reading of the manuscript.

**Conflicts of Interest:** The authors declare no conflict of interest.